\documentclass[conference]{IEEEtran}
\IEEEoverridecommandlockouts
\usepackage{cite}
\usepackage{amsmath,amssymb,amsfonts}
\usepackage{algorithmic}
\usepackage{enumerate}
\usepackage{graphicx}
\usepackage{textcomp}
\usepackage{pifont}
\usepackage[linesnumbered,ruled,noend]{algorithm2e}
\usepackage{listings}
\usepackage{color}
\usepackage{soul}
\usepackage{xcolor}
\usepackage{threeparttable}
\usepackage{booktabs}
\usepackage{makecell}
\usepackage[most]{tcolorbox}
\usepackage{colortbl}
\usepackage{subfigure}
\usepackage{rotating}
\usepackage{multirow}
\usepackage{hyperref}
\usepackage{enumitem}
\usepackage{balance}
\usepackage{pgfplots}
\usepackage{float}

\pgfplotsset{width=8cm,compat=1.9}
\SetCommentSty{\footnotesize\hspace{-3mm}}

\hypersetup{
  colorlinks,
  linkcolor=red, % Set link color
  urlcolor=blue,
  citecolor=black,
  bookmarksnumbered,
  unicode
}

\def\BibTeX{{\rm B\kern-.05em{\sc i\kern-.025em b}\kern-.08em
    T\kern-.1667em\lower.7ex\hbox{E}\kern-.125emX}}

\begin{document}

\title{Boosting Redundancy-based Automated Program Repair by Fine-grained Pattern Mining}

\author{
\IEEEauthorblockN{
Jiajun Jiang\IEEEauthorrefmark{1},
Fengjie Li\IEEEauthorrefmark{1},
Zijie Zhao\IEEEauthorrefmark{2}\IEEEauthorrefmark{7},
Zhirui Ye\IEEEauthorrefmark{3}\IEEEauthorrefmark{7}\thanks{\IEEEauthorrefmark{7}At the time of submission, Zijie Zhao and Zhirui Ye were undergraduate student at Tianjin University, Tianjin, China},
Mengjiao Liu\IEEEauthorrefmark{1},
Bo Wang\IEEEauthorrefmark{4},\\
Hongyu Zhang\IEEEauthorrefmark{5},
Junjie Chen\IEEEauthorrefmark{1}\IEEEauthorrefmark{6}\thanks{\IEEEauthorrefmark{6}Junjie Chen is the corresponding author for this work.}
}

\IEEEauthorblockA{\IEEEauthorrefmark{1}College of Intelligence and Computing, Tianjin University, Tianjin, China}

\IEEEauthorblockA{\IEEEauthorrefmark{2}Computer and Information Science Department, University of Pennsylvania, Philadelphia, USA}

\IEEEauthorblockA{\IEEEauthorrefmark{3}School of Engineering, Westlake University, Zhejiang, China}

\IEEEauthorblockA{\IEEEauthorrefmark{4}School of Computer and Information Technology, Beijing Jiaotong University, Beijing, China}

\IEEEauthorblockA{\IEEEauthorrefmark{5}School of Big Data and Software Engineering, Chongqing University, Chongqing, China}

\IEEEauthorblockA{
\{jiangjiajun, fengjie, mengjiaoliu, junjiechen\}@tju.edu.cn, bytez@cis.upenn.edu, yezhirui@westlake.edu.cn
}
\IEEEauthorblockA{
wangbo\_cs@bjtu.edu.cn, hyzhang@cqu.edu.cn
}
}

\newcommand{\plink}[1]{\textbf{\url{https://github.com/Feng-Jay/Repatt}}}

\newcommand{\jun}[1]{{\color{cyan}\ding{46}[Jiang: #1]}}

\newcommand{\bo}[1]{{\color{cyan}\ding{46}[Wang: #1]}}

\newcommand{\jcom}[1]{{\color{red}\ding{46}[Jiang: #1]}}

\newcommand{\tool}[1]{\textsc{Repatt}}

\newcommand{\combine}[1]{\textsc{Combine}}

\newcommand{\ye}[1]{{\color{magenta}\ding{46}[Ye: #1]}}

\newcommand{\z}[1]{{\color{purple}\ding{46}[Comment: #1]}}

\newcommand{\fengjie}[1]{{\color{brown}\ding{46}[fengjie: #1]}}

\newcommand{\hy}[1]{{\color{red}\ding{46}[HY: #1]}}

\newcommand{\add}[1]{{\color{blue}#1}}
\newcommand{\del}[1]{{\color{red}\sout{#1}}}
\renewcommand{\add}[1]{#1}
\renewcommand{\del}[1]{}

\newcommand{\cradd}[1]{{\color{blue}#1}}
\newcommand{\crdel}[1]{{\color{red}\sout{#1}}}
\renewcommand{\cradd}[1]{#1}
\renewcommand{\crdel}[1]{}

\definecolor{lightgray}{gray}{0.9}
\definecolor{dkgreen}{rgb}{0,0.6,0}
\definecolor{gray}{rgb}{0.5,0.5,0.5}
\definecolor{mauve}{rgb}{0.58,0,0.82}
\newcommand{\highlight}{\cellcolor{grey}}

\newcommand\mycommfont[1]{\scriptsize\ttfamily\textcolor{blue}{#1}}
\SetCommentSty{mycommfont}

\newtheorem{definition}{Definition}

\lstdefinestyle{Java}{ % Define a style for your code snippet, multiple definitions can be made if, for example, you wish to insert multiple code snippets using different programming languages into one document
	%    backgroundcolor=\color{highlight}, % Set the background color for the snippet - useful for highlighting
	language=Java,
	basicstyle=\scriptsize\ttfamily, % The default font size and style of the code
	breakatwhitespace=false, % If true, only allows line breaks at white space
	breaklines=true, % Automatic line breaking (prevents code from protruding outside the box)
	captionpos=b, % Sets the caption position: b for bottom; t for top
	commentstyle=\color[rgb]{0.0, 0.5, 0.69},%\color[rgb]{0,0.6,0}, % Style of comments within the code - dark green courier font
	deletekeywords={}, % If you want to delete any keywords from the current language separate them by commas
	%escapeinside={\%}, % This allows you to escape to LaTeX using the character in the bracket
	escapeinside={<@}{@>},
	firstnumber=1, % Line numbers begin at line 1
	frame=lines, % Frame around the code box, value can be: none, leftline, topline, bottomline, lines, single, shadowbox
	frameround=tttt, % Rounds the corners of the frame for the top left, top right, bottom left and bottom right positions
	keywordstyle={[1]\color{blue!90!black}},
	keywordstyle={[3]\color{red!80!orange}},
	morekeywords={String,int}, % Add any functions no included by default here separated by commas
	numbers=none, % Location of line numbers, can take the values of: none, left, right
	numbersep=-8pt, % Distance of line numbers from the code box
	numberstyle=\tiny\color[rgb]{0.1,0.1,0.1}, % Style used for line numbers
	rulecolor=\color{black}, % Frame border color
	showstringspaces=false, % Don't put marks in string spaces
	showtabs=false, % Display tabs in the code as lines
	stepnumber=1, % The step distance between line numbers, i.e. how often will lines be numbered
	stringstyle=\color[rgb]{0.58,0,0.82},
	tabsize=2, % Number of spaces per tab in the code
	backgroundcolor=\color{white}
}

\newcommand{\codeIn}[1]{{\ttfamily #1}}

\newcommand{\lin}[1]{{\scriptsize \textcolor{darkgray}{#1}}}

\newcommand{\hlc}[2]{{\setlength\fboxsep{0pt}\hspace{-3pt}\colorbox{#1} 
		{\begin{minipage}{\dimexpr\columnwidth-1\fboxsep+0pt\relax}
				\codeIn{\strut\hspace{3pt}#2}
			\end{minipage}}}}

% \setlength{\intextsep}{0.1\baselineskip plus 0.2\baselineskip minus 0.1\baselineskip}
% \setlength{\abovecaptionskip}{0.2\baselineskip}
% \setlength{\belowcaptionskip}{0.05\baselineskip}
% \setlength{\textfloatsep}{0.1\baselineskip plus 0.2\baselineskip minus 0.1\baselineskip}

% \newcommand{\distance}{1pt}
% \setlength{\textfloatsep}{\distance}%set distance between figure/tables on the top/bottom with text
% \setlength{\floatsep}{\distance}%set distance between figures or tables
% % \setlength{\intextsep}{\distance}%set distance between figures/tables in text with text
% \setlength{\dbltextfloatsep}{\distance} %distance between a figure/table spanning both columns and the text;
% \setlength{\dblfloatsep}{\distance} %distance between two figures/tables spanning both columns.

\maketitle

\begin{abstract}
%Automated program repair (APR) has the potential to reduce the debugging overhead of human developers and thus has been widely studied in the last decade. Among existing studies, r
Redundancy-based automated program repair (APR), which generates patches by referencing existing source code, has gained much attention since they are effective in repairing real-world bugs %and have 
with good interpretability.
%gained much attention because of their better interpretability. 
However, since existing approaches either demand the existence of multi-line similar code or %generate patches by randomly referencing existing code, 
randomly reference existing code, they can only repair a small number of bugs with many incorrect patches, hindering their wide application in practice. In this work, we aim to improve the effectiveness of redundancy-based APRs by exploring more effective source code reuse methods for improving the number of correct patches and reducing incorrect patches. Specifically, we have proposed a new repair technique named \tool{}, which incorporates a two-level pattern mining process for guiding effective patch generation (i.e., token and expression levels). We have conducted an extensive experiment on the widely-used Defects4J benchmark and compared \tool{} with \crdel{nine}\cradd{ten} state-of-the-art APR approaches. The results show that 
% \fengjie{although our approach did not outperform all baselines considering all comparison aspects,} 
it complements existing approaches by repairing {9} unique bugs compared with the latest Large Language Model (LLM)-based and deep learning-based methods and {19} unique bugs compared with traditional repair methods when providing the perfect fault localization. 
In addition, when the perfect fault localization is unknown in real practice, \tool{} significantly outperforms the baseline approaches by achieving much higher patch precision, i.e., {83.8\%}, although it repairs fewer bugs.
Moreover, we further proposed an effective patch ranking strategy for combining the strength of \tool{} and the baseline methods. The result shows that it repairs 124 bugs when only considering the Top-1 patches and improves the best-performing repair method by repairing 39 more bugs.
%\z{We also proposed a new approach that enables us to combine existing APR tools while preventing plausible patches from being provided to developers. Through this approach, \tool{} can provide correct patches for 124 projects in the Defects4J benchmark, improving the best SOTA works by 16 bugs.}
The results demonstrate the effectiveness of our approach for practical use.
\end{abstract}


\section{Introduction}
\label{sec:intro}

% \IEEEPARstart{A}{utomated}
%\jun{Redundancy-based APR is unclear.}
% \z{  \- The terms "online" and "offline" used for "token-level" mining and "expression-level" search aren't accurate IMHO.
    % \- Although the token-level prefix trees are built "offline", they are still searched for patch candidates when a bug is given for repair and thus used "online".
    % \- The expression-level search needs to construct S-TAC for each code block, which can be done "online" as described in the paper; however, the S-TACs for each program can be constructed "offline" too, just like the token-level prefix trees, to make the online search more efficient. It's unclear why the expression-level search component is totally "online".
    % \- Also, "token-level" vs. "expression-level" aren't clear either; it seems to be more accurate if renamed as single-line level vs. multi-line level based on the paper's description.}
    Automated program repair (APR) techniques have the potential to significantly reduce the debugging overhead of human developers and thus have been widely studied in the last decades. To date, a large number of APR approaches have been proposed and developed~\cite{Wen2018ContextAwarePG,xiong-icse17,xuan2016nopol,liu2019tbar,tananti,GenProg,long2016automatic,ISSTA18-SimFix,ssFix,SemFix,B.Le2016}. A typical APR approach takes a faulty program and a set of test cases, where at least one triggers the bug in the program, as inputs, and produces one or more (if possible) \textit{plausible} patches that can make all the test cases pass. The repair process can be typically viewed as a search problem, where the search space is all possible programs. Therefore, the core challenge for APR techniques is how to effectively and efficiently locate the desired patches within limited computing resources and time budget. To facilitate this process, existing techniques have employed multiple data sources, e.g., similar code~\cite{ISSTA18-SimFix,ssFix}, historical patches~\cite{Xuan2016History,le2017s3}, repair templates~\cite{liu2019tbar,PAR,ELIXIR}, etc., and leveraged diverse search algorithms, e.g., random search~\cite{RSRepair}, genetic programming~\cite{GenProg,GenProgTSE}, and advanced Large Language Model (LLM) and deep learning~\cite{zhu2021syntax,ye2022selfapr,jiang2023impact,jiang2021cure, xia2022less, wei2023copiloting, zhang2023gamma}, aiming to effectively refine the search space and thus improve the repair ability and efficiency. 
    
    Among existing APR techniques, redundancy-based~\cite{chen2018essence,white2019sorting,yang2021accelerating,chen2018remarkable} techniques have attracted much attention and achieved considerable success. These techniques are on the basis of the \textbf{plastic surgery hypothesis} that ``\textit{changes to a codebase contain snippets that already exist in the codebase at the time of the change, and these snippets can be efficiently found and exploited}~\cite{DBLP:conf/sigsoft/BarrBDHS14}''. In other words, the code snippets for patch generation can be found within the buggy projects themselves. Therefore, redundancy-based APR approaches leverage the plastic surgery hypothesis to generate patches by searching and referencing existing code. For example, GenProg~\cite{GenProg,GenProgTSE}, the pioneer of modern APR techniques, generates patches by directly reusing existing code under the guidance of genetic programming algorithms. Although the recent advance of LLMs has shown the promise to repair more bugs~\cite{jiang2023impact, xia2022less, wei2023copiloting, zhang2023gamma}, they still suffer from the interpretability issue due to the complexity and the random nature of deep neural networks as they are typically used in a  black-box fashion. As reported by previous studies~\cite{liuonsystematic,Motwani2022Quality,wang2019how}, the patches generated by SimFix, a representative redundancy-based APR approach, are of high quality under various testing scenarios. In addition, redundancy-based  techniques also complement those learning-based techniques~\cite{zhu2021syntax,ye2022selfapr,jiang2023impact} (our evaluation results in Section~\ref{sec:result} further prove it). Additionally, redundancy-based techniques tend to have better interpretability as they are on the basis of the \textit{plastic surgery hypothesis}, and the patches are typically generated by referring human-written code snippets. In particular, they can repair a portion of relatively complex bugs~\cite{Yang2023transplantFix,ISSTA18-SimFix,ssFix}.

Although existing redundancy-based techniques are demonstrated to be effective for repairing real-world bugs, they still face 
%However, reusing existing source code for patch generation has 
two major challenges: (1) \textit{How to effectively and efficiently locate the reference code elements among a large-scale codebase?} Existing approaches identify similar snippets using structural~\cite{ISSTA18-SimFix,Yang2023transplantFix} or token-based~\cite{ssFix} features, assuming similarity implies functional equivalence. However, as shown in our preliminary study (Section~\ref{sec:preliminary}), coarse-grained similar snippets are rare in practice, limiting their usability. (2) \textit{How to reuse existing code for constructing new patches?} Existing approaches reuse code by comparing the difference between faulty and reference code via abstract syntax trees~\cite{ISSTA18-SimFix,ssFix} or program dependency graphs~\cite{Yang2023transplantFix}. 
However, code elements under different contexts tend to vary greatly in structures~\cite{rattan2013software}, limiting the effectiveness of existing techniques, i.e., repairing a small number of real-world bugs with many incorrect patches. 

As explained above, existing redundancy-based approaches reuse code either completely randomly (e.g., RSRepair~\cite{RSRepair} and GenProg~\cite{GenProg}) or depending on the existence of similar code snippets of multiple lines (e.g., HDRepair~\cite{7476644}, SimFix~\cite{ISSTA18-SimFix} and TransplantFix~\cite{Yang2023transplantFix}). 
However, our preliminary study reveals that most reusable elements appear at a finer-grained level—e.g., 89.3\% contain fewer than three tokens (Section~\ref{sec:preliminary}). This suggests that finer-grained code reuse is crucial.
% As it will be presented in our preliminary study, though the reusable code elements for patch generation may appear at different granularities, most of them actually exist at the finer-grained level, e.g., 89.3\% reusable code elements contain less than three tokens (Section~\ref{sec:preliminary}).
% In contrast, existing approaches usually reuse them at a fixed coarse-grained granularity (i.e., depending similar code hunks) and thus lack of flexibility.
%the reusable code elements for patch generation may appear at different granularities, e.g., token levels, statement levels, or even block levels,  while existing approaches usually reuse them at a fixed granularity and thus lack of flexibility.
In this paper, we investigate the possibility of improving redundancy-based APR techniques by reusing code elements at a finer-grained level. Specifically, the basic idea of our approach is that \textbf{fine-grained code elements related to similar code semantics tend to co-appear nearby in the program}, which has been well-studied by existing research~\cite{Hindle2012,ray2016naturalness,khanfir2022codebert,tu2014localness}. \del{We call this phenomenon \textit{locality property} of source code.}This phenomenon, which we refer to as the \textit{locality property} of source code, suggests that semantically related code components—such as variables, functions, or expressions—are often found in close proximity within the same code block or module. % (a.k.a., \textit{code naturalness}).
%in our preliminary study, we find that the code elements for patch generation also tend to co-appear in other locations, we call this phenomenon as the \textit{locality property} of source code. 
Based on this property, we proposed a new APR approach that 
%by identifying the usage patterns of fine-grained code elements, which can 
 effectively guides patch generation by identifying the usage patterns of fine-grained code elements for confining the patch space and thus overcomes the first challenge of redundancy-based APR.  To overcome the second challenge, i.e., reusing existing code under different contexts,
 % for constructing patches, 
 we designed a new code representation method, i.e., S-TAC, which decomposes complex code expressions and statements into a unified form by ignoring context-specific features (e.g., code structures and operators). It can reduce the negative impact of code structures on code matching during patch generation but still preserve the \textit{locality property} of source code.

To evaluate the performance of our approach, we have implemented it as an APR tool named \tool{} and conducted an extensive study by comparing it with \crdel{nine}\cradd{ten} state-of-the-art APR approaches, including \crdel{three}\cradd{four} best-performing traditional repair techniques and six latest LLM-based and deep learning-based techniques. The experimental results show that although our approach did not outperform all baselines considering all comparison aspects, it complements existing approaches by correctly repairing many unique bugs.
% that cannot be repaired by the baseline approaches. 
Specifically,
\add{\tool{} repairs 19 unique bugs that cannot be repaired by traditional methods and 9 unique bugs that neither LLM-based nor deep learning-based methods can repair.}
% \tool{} repairs 19 and 9 unique bugs that cannot be repaired by the traditional and deep learning-based methods, respectively.
In particular, 5 bugs repaired by \tool{} have never been repaired by all the baselines. Additionally, when the perfect fault localization is unknown, which is a more realistic scenario, \tool{} can significantly outperform the baseline approaches by achieving 15.6\%-51.7\% higher patch precision, which is critical for practical use of APRs. 
Additionally, we also made the first attempt to combine the strength of \tool{} and multiple existing approaches by further designing a patch ranking strategy. The evaluation result shows that the combined method can effectively repair 124 bugs when only considering Top-1 patches, 39 more bugs than the best-performing method (i.e., TBar). Our results demonstrate that our approach is indeed effective in improving the performance of existing APR techniques. 

In summary, we make the following major contributions.

\begin{itemize}
	\item A new redundancy-based APR technique that can flexibly reuse source code at a finer granularities.
	\item A novel code representation method, which overcomes the diversity of code contexts, for better code search and reuse.
	\item An extensive study to evaluate the performance of our approach by comparing it with state-of-the-art approaches.
	\item The first attempt to combine the strength of multiple APR approaches, revealing its promise for practical use.
	\item We make all our experimental results and implementations publicly available to facilitate replication and comparison~\cite{homepage}. 
 
% \href{https://anonymous.4open.science/r/RepattTool}{\textbf{https://anonymous.4open.science/r/RepattTool}}
 % \plink{}
\end{itemize}

\section{Motivation}
\label{sec:motivation}
% In this section, we present a preliminary study and leverage examples to motivate the feasibility of our approach.

\subsection{Preliminary Study}
\label{sec:preliminary}

%\begin{figure}[t]
%	\centering
%	\includegraphics[width=0.99\columnwidth]{fig/newtoken.pdf}
%	\caption{Ratio of required tokens that co-appear with some existing tokens in the faulty line to all required tokens for repairing those bugs. Statistics on 125 bugs that require one-line code change from the Defects4J benchmark.}
%	\label{fig:tokenratio}
%\end{figure}

% \begin{figure*}[t]
% 	\centering
% 	\includegraphics[width=0.99\textwidth]{fig/frag_length.png}
% 	\caption{The number of tokens in code elements that can be found in the faulty program for repairing corresponding bugs.\jcom{@Zijie, Similarly, change ``jacksoncore'' to ``JacksonCore'', make the figure clearer (higher resolution). I cannot see the x/y-labels clear.}}
% 	\label{fig:frag_len}
% \end{figure*}
\begin{figure}[tb]
    \centering
     \includegraphics[width=0.98\columnwidth]{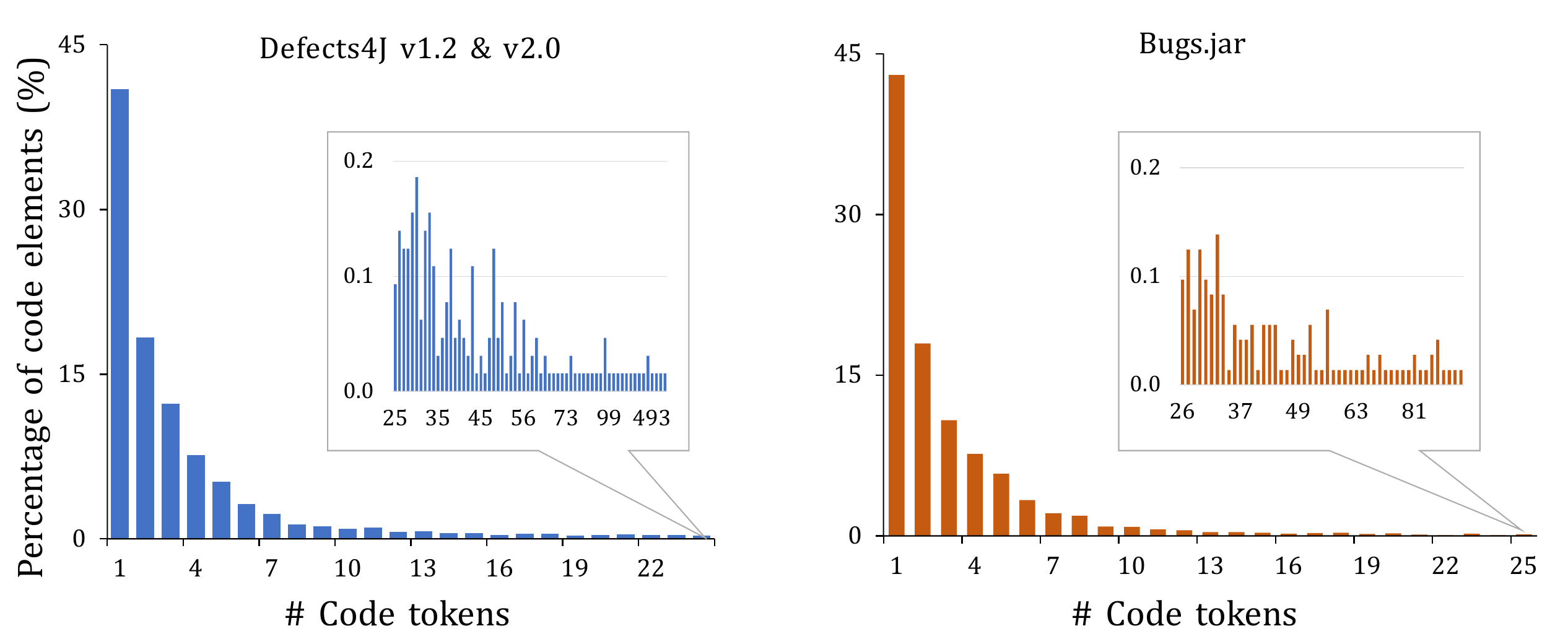}
%    \subfigure{ \includegraphics[width=0.45\textwidth]{fig/d4j_plot.pdf}
%    }
%    \subfigure{ \includegraphics[width=0.45\textwidth]{fig/bugsdotjar_plot.pdf}
%    }
    % \Description{The length distribution of code elements that can be found in the faulty program for repairing related bugs.}
    \caption{The length distribution of code elements that can be found in the faulty program for repairing 
    % related bugs based on the 
    564 and 609 bugs from benchmarks Defects4J and Bugs.jar.}
\label{fig:frag_len}
\vspace{-10pt}
\end{figure}

As previously mentioned, no existing studies have explored the potential of searching for and reusing finer-grained code elements for patch generation. To investigate the feasibility and necessity of this approach in APR, we conducted a preliminary study following the idea proposed by Barr et al.~\cite{DBLP:conf/sigsoft/BarrBDHS14}.

    Specifically, our study aims to understand the granularity of reusable code elements for patch generation. Given a bug and its associated patch, we first extracted newly added code elements from the patch and searched for them in the faulty program, starting from coarse-grained elements and gradually decomposing them into finer-grained components based on the syntax tree structure. That is, if a code element (e.g., ``\codeIn{a + f(b,c)}'') was not found, we would further decompose it into finer-grained code elements of the code (e.g., ``\codeIn{a}'', ``\codeIn{+}'', and ``\codeIn{f(b,c)}''). This process iterated until no further decomposition was possible. Finally, we analyzed the granularity of code elements that can be found in faulty programs. In particular, since AST structures cannot directly reflect granularity (e.g., an \codeIn{expression} in the AST can be either a variable like ``\codeIn{a}'' or a complex expression like ``\codeIn{f(a, b)}''.), we measured granularity by the number of tokens in each code element.
 
	We conducted our empirical study on two widely-used datasets of real-world bugs, i.e., Defects4J~\cite{just2014defects4j} and Bugs.jar~\cite{Saha2018bugs}. In particular, we filtered out bugs whose patches involve more than one Java file since they are still hard to repair at present~\cite{white2019sorting, ISSTA18-SimFix,xiong-icse17,xia2023automated,jiang2023impact,zhu2021syntax,chen2019sequencer,li2020dlfix}, resulting in 564 bugs from Defects4J and 609 bugs from Bugs.jar across 24 projects.
    Figure~\ref{fig:frag_len} presents the statistical results. The \textit{x-axis} denotes the number of tokens in a code element and the \textit{y-axis} denotes the percentages of code elements that can be reused for patch generation. The results show on average more than 89.3\% code elements contain less than three tokens, and 42\% elements only contain one token. In contrast, only a very limited portion of code elements exist at a coarse-grained granularity (i.e., more than three tokens). In particular, this is not a special case but holds over different programs in our study. The result indicates that the reusable code elements indeed exist at a fine-grained granularity, which will significantly limit the effectiveness of existing redundancy-based APR approaches. Moreover, it also reflects that designing APR approaches by reusing finer-grained code elements is promising. 

    % \add{Additionally, we analyzed the distribution of the number of tokens in a code element for each project, as shown in Figure~\ref{fig:frag_proj}. In this figure, the \textit{x-axis} represents different projects from each dataset, and the \textit{y-axis} represents the number of tokens in a code element. Please note that each project may have multiple patches and a patch may involve multiple code elements. The results indicate that each project follows the conclusion shown in Figure~\ref{fig:frag_len}, where most code elements contain no more than three tokens.}

% \begin{figure}[htb]
%     \centering
%      \includegraphics[width=0.98\columnwidth]{fig/preliminary2.pdf}
%     \caption{\add{The distribution of code element lengths found in faulty programs for repairing 564 and 609 bugs from the Defects4J and Bugs.jar benchmarks across different projects.}}
% \label{fig:frag_proj}
% \end{figure}
 
	However, achieving accurate and fine-grained code reuse for patch generation is challenging. The reasons are twofold: (1) It will significantly enlarge the search space of candidate patches, making it harder to efficiently and correctly find the desired code elements. (2) Due to weak test suites~\cite{PatchPlausibility,xin2017identifying,xiong-icse18}, plausible (i.e., can pass all the test cases) but incorrect patches are more likely to be generated, increasing the manual effort required for validation and significantly affect the practical usability of APR approaches.
    % Plausible (i.e., can pass all the test cases) but incorrect patches will be more easily generated due to the problem of weak test suite~\cite{PatchPlausibility,xin2017identifying,xiong-icse18}. It will cause heavier manual effort for patch validation and thus significantly affect the usability of the APR approach in practice. 
    To overcome these challenges, we propose an effective APR approach that efficiently locates the desired code elements for patch generation leveraging the \textit{locality property} of source code as its core idea. The details will be introduced in Section~\ref{sec:app}.

	\subsection{Running Example}
	\label{sec:example}
	
	% Motivated by our preliminary study, reusing finer-grained code elements can enhance redundancy-based program repair, making it both feasible and promising. 
    In this section, we use real-world bug examples to illustrate how the \textit{locality property} of source code can guide patch generation.
	Listing~\ref{lst:codec3} presents the patch code of Codec-3 from the Defects4J~\cite{just2014defects4j}, where a line starting with ``+'' denotes newly added code while starting with ``-'' denotes deleted code.
    
%   \begin{figure}[htb]
%     \centering
%      \includegraphics[width=0.7\columnwidth]{fig/test.pdf}
%     \caption{The constant token ``3'' usually co-appears with tokens ``value'' and ``index''}
% \label{fig:expr_examples}
% \end{figure}

	In this example, the constant parameter ``\codeIn{4}'' was mistakenly used at line 455, and the desired parameter is ``\codeIn{3}''. This bug cannot be fixed by existing redundancy-based APR techniques due to either the large search space (e.g., GenProg~\cite{GenProg,GenProgTSE}, RSRepair~\cite{RSRepair}, etc.) or the nonexistence of multi-line similar code snippets for reference (e.g., SimFix~\cite{ISSTA18-SimFix}, TransplantFix~\cite{Yang2023transplantFix}, etc.). However, when it comes to the fine-grained token level, we will find that the constant number ``3'' usually co-appears with the tokens ``value'' and ``index'' elsewhere in the program.
    % \add{as shown in Figure~\ref{fig:expr_examples}}. 
    These tokens exhibit a strong correlation (i.e., the \textit{locality property} of source code). Leveraging this property makes it both feasible and efficient to locate the correct code elements for patch generation.
        \begin{lstlisting}[ style=Java,caption=Patch code of Codec-3 in Defects4J,label=lst:codec3] 
<@\lin{452}@>    if ((contains(value, 0 ,4, "VAN ", "VON ") || ...)){
<@\lin{453}@>        //-- obvious germanic --//;
<@\lin{454}@>        result.append('K');
<@\hlc{red!15}{\lin{455} - \quad\} else if(contains(value, index + 1, 4, "IER")) \{}@>
<@\hlc{green!15}{\qquad   + \quad\} else if(contains(value, index + 1, 3, "IER")) \{}@>
<@\lin{456}@>      result.append('J');
<@\lin{457}@>    } else {...
\end{lstlisting}
% \vspace{-8pt}
        % \noindent\begin{minipage}{\linewidth}
\begin{lstlisting}[ style=Java,caption=Patch code of Gson-17 in Defects4J,label=lst:gson17] 
<@\lin{98}@>    public Date read(JsonReader in) throws IOException {
<@\hlc{red!15}{\lin{99}\quad  -\qquad   if (in.peek() != JsonToken.STRING) \{}    @>
<@\hlc{red!15}{\lin{100} -  \qquad\qquad throw new JsonParseException("The date ...");}@>
<@\hlc{green!15}{\qquad +\qquad   if (in.peek() == JsonToken.NULL) \{ }@>
<@\hlc{green!15}{\qquad + \qquad\qquad  in.nextNull(); }@>
<@\hlc{green!15}{\qquad +  \qquad\qquad   return null; }@>
<@\lin{101}@>     }
<@\lin{102}@>     Date date = deserializeToDate(in.nextString()); ...
\end{lstlisting}
% \end{minipage}
\vspace{-8pt}
	\begin{lstlisting}[ style=Java,caption=The reference code for repairing Gson-17,label=lst:gson17donor] 
<@\lin{1}@>    if (in.peek() == JsonToken.NULL) {
<@\lin{2}@>        in.nextNull();
<@\lin{3}@>        return null;
<@\lin{4}@>    }
<@\lin{5}@>    in.beginArray();
\end{lstlisting}
%<@\lin{6}@>  while (in.hasNext()) {
%<@\lin{7}@>      E instance=componentTypeAdapter.read(in);
%<@\lin{8}@>      list.add(instance);
%<@\lin{9}@>  }
% \vspace{-8pt}
        % \input{code/gson15}
        % \input{code/gson15dornor}
	In fact, the \textit{locality property} of source code may not only exist at the token level but may also exist at higher levels, e.g., expression or statement levels. For example, Listing~\ref{lst:gson17} presents the patch code of Gson-17 from the Defects4J benchmark, while Listing~\ref{lst:gson17donor} presents a reference code that can be used for patch generation. According to the reference code, the expression ``\codeIn{in.peek()}'' may be correlated with the expressions ``\codeIn{JsonToken.NULL}'', ``\codeIn{in.nextNull()}'' and ``\codeIn{return null}''.
    It can be seen that, although the reference code can provide the required code elements for generating the desired patch in these examples, how to utilize them is still challenging since the possible combinations of code changes according to the reference code are still too many,
    % next line replace to example gson15
    such as replacing the \codeIn{throw} statement with ``\codeIn{in.nextNull()}'' or inserting ``\codeIn{in.nextNull}''.
    % such as replacing the \codeIn{IllegalArgumentException} object with ``\codeIn{NumberFormatExceptio}'' or inserting ``\codeIn{!lenient()}''.
 % as a new statement. 
 In addition, replacing the conditional expression
% ``\codeIn{JsonToken.STRING}'' with ``\codeIn{JsonToken.NULL}'' 
 also does not fix the bug due to the incorrect operator ``\codeIn{!=}''. In particular, the code structures of reference code and the faulty code may also be different in practice, making patch generation harder. To address this challenge, we propose a novel code representation method that decomposes complex expressions into a \textit{unified} simple form, which ignores operators and complex code structures, e.g., \codeIn{!=}, \codeIn{for}, \codeIn{while}, etc., but still preserves the \textit{locality property} of source code. 
\section{Framework}
\label{sec:app}

\begin{figure*}[t]
	\centering
	\includegraphics[width=0.98\textwidth]{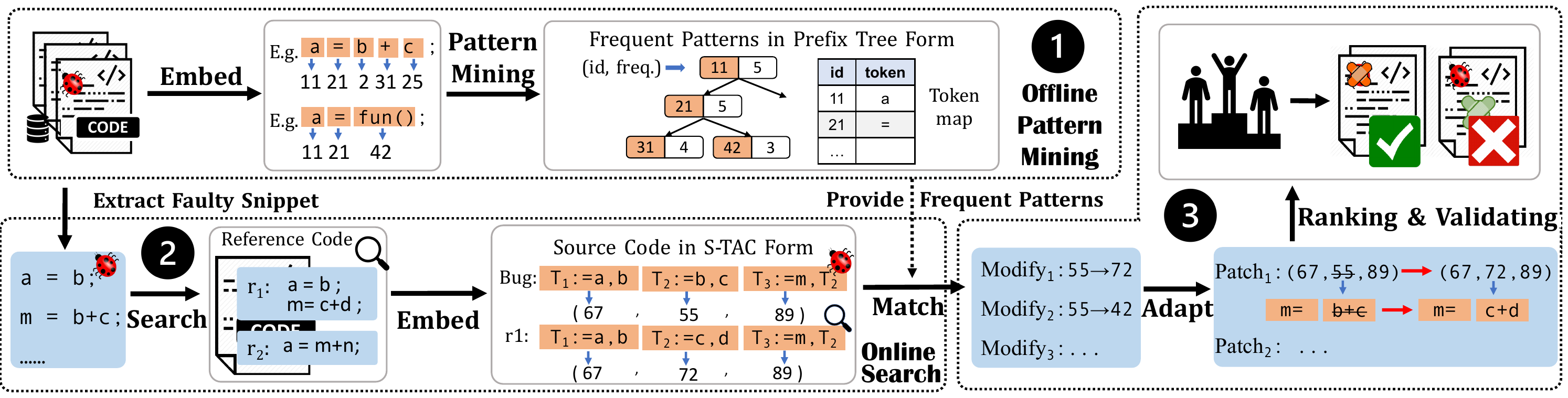}
        % \Description{Overview of our approach \tool{}}
	\caption{Overview of our approach \tool{}. Given a buggy project, \ding{182} \tool{} first mines a set of token sequence patterns and represents them as prefix trees. \ding{183} Then, based on the faulty code snippet, \tool{} online searches a set of similar code snippets and transforms them into S-TAC form as reference patterns. \ding{184} Finally, \tool{} generates candidate patches by matching the faulty code with the reference patterns and validates them one by one.}
	\label{fig:overview}
    \vspace{-10pt}
\end{figure*}
% \vspace{-10pt}

In this section, we introduce the details of our approach (named \tool{}). Figure~\ref{fig:overview} presents the overview of \tool{}. In general, \tool{} generates candidate patches by leveraging the 
% \del{\textit{plastic hypothesis}}
{\textit{plastic surgery hypothesis}} and the \textit{locality property} of source code, utilizing both token-level pattern mining and expression-level code search to handle different granularities of code changes. Token-level pattern mining, an \textit{offline} process, constructs a query-efficient code pattern database for repairing single-line bugs (\textit{ref.} Listing~\ref{lst:codec3}). In contrast, expression-level code search, an \textit{online} process, identifies reference code elements for multi-line fixes (\textit{ref.} Listing~\ref{lst:gson17}). Their differing strategies stem from (1) token-level mining focusing on small-scale patterns within single lines, while expression-level patterns span multiple lines, significantly expanding the search space, and (2) token-level patterns often introducing numerous small changes, impacting efficiency, while the expression-level patterns correlated to a certain location (i.e., the faulty code) will be very limited (refer to Figure~\ref{fig:frag_len}). Therefore, we confine the token-level patterns to those that are frequent to balance efficiency.

\subsection{Offline Pattern Mining}
\label{subsec:mining}

Based on our preliminary study (Section~\ref{sec:preliminary}), most reusable code elements for patch generation exist at a fine-grained level, typically one or two tokens. To leverage these elements, \tool{} employs a novel token-level pattern mining algorithm guided by the \textit{locality property} of source code. This process is conducted offline over the given faulty project, constructing a database of token usage patterns to support efficient online patch generation. As aforementioned, we only consider the token patterns within single lines to confine the search space and avoid involving too much noise. In particular, it is common that the tokens used in different code lines share both commonalities and diversities. For example, the token sequence in some code lines can be (``a'', ``b'', ``c''), while it may also be (``a'', ``d'', ``c'') in some other code lines. To preserve the semantics of programs, we keep the token orders and design a \textit{skip-fashion} pattern mining algorithm to address the problem of diversity among code lines. Additionally, each token is assigned a unique ID for efficient comparison, while separators (e.g., ``,'' and ``('') and structural keywords (e.g., ``if'' and ``for'') are removed as they do not contribute to token-level patch generation.

% \add{Unlike existing pattern mining approaches~\cite{meng2011sydit, jiang2019inferring, nguyen2019graph, bavishi2019phoenix} that retain multi-line code structure and concrete semantic information,} 

%Algorithm~\ref{alg:mining} presents the details of the mining process.

%It's common for developers to use incorrect variables or functions. However, due to the many available variables and functions, the search space is huge. Providing reference codes for repair can effectively reduce search space. But for the token level, the method of online search is not working. First, due to the lack of information, it is difficult to compare whether two small code fragments are similar. What's more, the available variables and their combinations can lead to a still large search space. To overcome these challenges, we construct the offline code base to mine the reference code.

% Before introducing the details, we first formally define some notions that will be used later. 
Specifically, to improve the pattern mining process, we design a data structure -- \textit{Prefix Embedding Tree}, which is defined as follows.

\begin{definition}
    \textbf{(Prefix Embedding Tree)}: a prefix embedding tree constitutes a set of nodes, each of which is a tuple of $t=\left\langle tok, id, p, C, sup\right\rangle$, where $id$ denotes a unique embedding of a certain token $tok$, $p$ denotes the parent node of $t$, $C$ denotes a set of child nodes of $t$, while $sup$ represents the frequency of the token sequence from the root to the current node $t$.
\end{definition}

Furthermore, we use $t.childNode(id)$ to obtain the child node of $t$ with the embedding $id$, and use $t.setChildNode(id, chd)$ to set the node $chd$ with embedding \textit{id} as the child node of $t$. In addition, we use \textit{findOrCreate(trees, tok)} to find the tree rooted at the token \textit{tok} in \textit{trees}.
%\jcom{@Zijie, I remember Algorithm~\ref{alg:mining} has been changed. Please update accordingly.}
%\z{Finished, need check}
% \z{It's unclear how the "findOrCreate(...)" function works in Algorithm 1.
% It's likely wrong that it says "...or create a new tree in case it does exist." It should be "...or create a new tree in case it does *not* exist."(FIXED)}
% \z{It seems a new tree would be created for each token in a given sequence according to the while loop at Line 10 in Algorithm 1; then, Figure 3 can be enriched better to reflect that, e.g., by adding another tree rooted at the "value" token.}
% \z{The skip-fashion would add some sub-sequences of tokens into the prefix trees, but it doesn't seem to consider syntactical constraints on the token sub-sequences, and thus those sub-sequences may not correspond to syntactical valid expressions in the original code. Could such code patterns be used for repair candidates too? (FIXED)}
According to these notions, we present the pattern mining process in Algorithm~\ref{alg:mining}.
%\jun{Prefix embedding tree is unclear, constructed for individual statement or the complete project, what about the overhead.}
In general, given the faulty program, \tool{} first decomposes all the code lines in the program into a set of token sequences (i.e., \textit{ES}s), where each token sequence corresponds to one line of code. Then, it constructs the prefix embedding trees (i.e., \textit{trees}), which stores the token patterns with corresponding usage frequencies.

% \vspace{-10pt}
\begin{algorithm}[h]
	\footnotesize
	\caption{Tree Building and Pattern Construction}
	\label{alg:mining}
	\SetKwInput{KwInput}{Input}                % Set the Input
	\SetKwInput{KwOutput}{Output}              % set the Output
	\DontPrintSemicolon
	
	\KwInput{\textit{ESs:} \textcolor{gray}{sequences of code element embeddings (IDs)
 % corresponding to all statements.
 }} 
	\KwOutput{\textit{trees:} \textcolor{gray}{frequent pattern in the form of prefix tree.}}
	%	\KwData{}
	
	% Set Function Names
	\SetKwFunction{FMain}{buildTree}
        \SetKwFunction{FStart}{build}
	\SetKwFunction{FBuild}{traverse}
	
	\SetKwProg{Fn}{Function}{:}{}

        \Fn{\FStart{\textit{ESs}}}{
            \textit{trees} $\leftarrow \emptyset$ \;
            \ForEach{$S$ in ESs}{ 
                \textit{updated} $\leftarrow \emptyset$ \tcp{Avoid counting duplicated tokens}
                \FMain{S, trees, updated} \tcp{Build trees}
            }
            \Return{trees} \tcp{Embedding trees in programs}
        }
	\Fn{\FMain{\textit{seq, trees, updated}}}{
		\textit{start} $\leftarrow$ 0\;
		\While{start $<$ len(seq)}{
			\textit{root} $\leftarrow$ \textit{findOrCreate(trees, seq[start])} \tcp{Find node}
			\textit{root.sup} $\leftarrow$ \textit{root.sup + 1} \tcp{Increase the frequency}
			\textit{start} $\leftarrow$ \textit{start + 1}\;
			\FBuild{root, seq[start:], 0, 1, updated}\;
			\textit{trees} $\leftarrow$ \textit{trees} $\cup$ \{\textit{root}\}\;
		}
	}
	\Fn{\FBuild{\textit{tree, seq, skip, length, updated}}}{
		\uIf{length $\geq$ MAX\_LEN or len(seq) $==$ 0}{
			\KwRet{}
		}
		\ElseIf{skip $<$ MAX\_SKIP} {
        % \tcp{Skip the token} 
				\FBuild{tree, seq[1:], skip+1, length, updated} 
			}
				\textit{node} $\leftarrow$ \textit{tree.childNode(seq[0])} \tcp{Not skip the token}
				\uIf{node is None}{
					\textit{node} $\leftarrow$ create a new node of id \textit{seq[0]}\;
					\textit{tree.setChildNode(seq[0], node)}\;
				}
				\uIf{node is not in updated}{
					\textit{node.sup} $\leftarrow$ \textit{node.sup + 1} \tcp{Increase frequency}
                        \textit{updated} $\leftarrow$ \textit{updated} $\cup$ \{\textit{node}\}\;
				}
				\FBuild{node, seq[1:], skip, length + 1, updated}\;
%			}
	}
\end{algorithm}

Specifically, for each token sequence $S\in$ \textit{ES}s (line 3), \tool{} either creates new trees using it or expands existing trees by updating pattern frequencies via \textit{buildTree(*)} (Line 5). In this process, \tool{} maintains a set \textit{updated} to record the nodes whose frequencies have been calculated for avoiding duplicated counting. More specifically, given the token sequence \textit{seq} and \textit{trees} that have already constructed, \tool{} performs a top-down tree building process, iterating over sub-sequences starting at different positions in \textit{seq} (Line 9). That is, for each token sub-sequence starting at \textit{start}, \tool{} finds the tree (or creates a new one) whose root node is associated with the token \textit{seq[start]} (Line 10) and updates its frequency (Line 11). The tree is then recursively expanded via \textit{traverse()} using a \textit{skip-fashion} pattern mining strategy, where a token can either be included (Lines 20-27) or skipped (Lines 18-19), constrained by MAX\_SKIP (Line 18). For each token, \tool{} locates an existing node (Line 20) or creates a new one if absent (Lines 21-23), updating its frequency as needed (Lines 24-26). A pattern is finalized upon reaching the maximum sequence length MAX\_LEN. Our evaluation further examines the impact of MAX\_LEN and MAX\_SKIP on \tool{}’s performance.

\begin{figure}[t]
	\centering
	\includegraphics[width=\columnwidth]{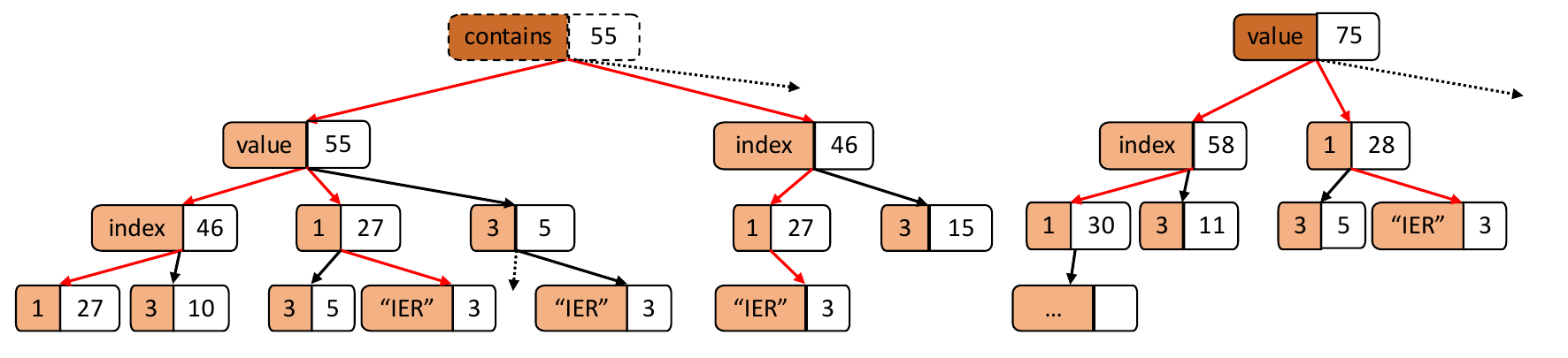}
        % \Description{\underline{Partial} prefix embedding trees. We use concrete tokens to represent the token ids for ease of understanding.}
	\caption{\underline{Partial} prefix embedding trees. We use concrete tokens to represent the token ids for ease of understanding.}
	\label{fig:tree}
    \vspace{-15pt}
\end{figure}

Taking the faulty code Codec-3 (Listing~\ref{lst:codec3}) as an example, the token sequence of the faulty code line will be $S$=(``contains'', ``value'', ``index'', ``+'', ``1'', ``4'', ``IER''), based on which the constructed prefix embedding trees are presented in Figure~\ref{fig:tree}, where we use the {\color{red} red line} to highlight the patterns related to the token sequence $S$. Consequently, the token sequence from the root node to the leaf node denotes a possible pattern and the number denotes its frequency. For example, the frequency of the pattern (``contains'', ``value'', ``1'', ``IER'') is 3. In particular, each distinct token will be the root node of an individual prefix embedding tree. For instance, we present two prefix embedding trees in Figure~\ref{fig:tree} that are related to the token sequence $S$. As a result, when given a buggy sequence, we can only search a very limited number of embedding trees whose root tokens are included in the sequence, which can promote the searching process. During pattern mining, patterns with a frequency below a predefined threshold MIN\_SUPPORT are discarded. A higher frequency suggests that a pattern is more common and thus more likely to be reusable for repair. In the repair phase, given a faulty code token sequence $S'$ , \tool{} searches for all matching patterns in the embedding trees and generates patches based on these references.

\subsection{Online Code Search and Representation}
\label{sec:online}
The expression-level code search aims to find code usage patterns across multi-lines of code for reference, which may cause a large search space if it is also considered like the token-level pattern mining since there may be hundreds or thousands of lines of code even in a single method. As a result, the expression-level pattern mining process is designed as an online code search process, which automatically identifies a set of reference code when given the faulty code. Following existing work~\cite{ISSTA18-SimFix}, \tool{} extracts no more than three lines of code respectively before and after the given faulty line as the \textit{faulty snippet}, and then finds the $N$ most similar snippets via measuring the code similarity between the faulty and reference snippets. Specifically, \tool{} extracts a vector from each snippet, where each element in the vector represents a feature and the feature value represents the number of corresponding AST node types in the snippet by following the study~\cite{ISSTA18-SimFix}. Then, we calculate the cosine similarity~\cite{Cosine} between two vectors for code ranking.

To adapt the reference code to the buggy code that many have complex contexts in real practice, we propose a new code representation method, named ``Simplified Three-Address Code'' (\textit{abbr.} S-TAC). It unifies the code representations by ignoring the code structures under different contexts but still preserves the \textit{locality property} of source code.

\begin{definition}
    \textbf{(Simplified Three-Address Code)}: an S-TAC is a triple of $\left\langle T, t_1, t_2\right \rangle$, where $T$ is an intermediate symbol to represent the expression correlated to $t_1$ and $t_2$, while
    $t_1$ and $t_2$ are two expressions that are either the intermediate symbols or simple items of variables, literal values, types, and function calls. In particular, $t_1$ and $t_2$ can be null if they are keywords, e.g., \codeIn{break} and \codeIn{return}.
    % \z{It's unclear if a S-TAC as defined on Page 5 allows at most two terms (without considering recursions) or an arbitrary number of terms (with recursions). Three-address code is named so in the literature because there could be at most three addresses in an expression; however, the definition of TAC in the paper doesn't follow this rule. This also causes my confusion on the example in Figure 4.}
\end{definition}
% According to this definition, the S-TAC representation can also be equivalently presented as follows. In particular, the constant symbol $T$ is not recursively defined. $T_i$ and $T_j$ are two constant symbols that are defined by a previous S-TAC and thus cannot be expanded further.
% \[
% \begin{array}{ccl}
%      T& :=& atom, atom 
%      ~|~ T_i, T_j \\
%      & | & T_i, atom
%      ~|~ atom, T_i\\
%      & | & T_i ~|~ atom\\
%      atom & := &variable~ | ~literal~|~ type ~| ~call | ~ field ~access
% \end{array}
% \]
% As we can observe, 

Additionally, we use $T:=t_1,t_2$ to represent $\left\langle T, t_1, t_2\right \rangle$ when there is no ambiguity.
Compared with traditional Three-Address Code (TAC)~\cite{aho2020compilers}, S-TAC ignores both the coarse-grained code structure information, e.g., \codeIn{if}, and the fine-grained operators, e.g., ``+''. The reason for ignoring the structure information is to make the source code from different contexts applicable for patch generation. For example, the conditional expression $a>b$ in \codeIn{if} statements can match that either in \codeIn{for} statements or in trinary conditional expressions, enabling a more flexible code match. Similarly, the fine-grained operators may also be specific to certain contexts and thus affect the matching and reuse of code.
In other words, our S-TAC representation can better reflect the \textit{locality property} of source code while reducing the noise induced by different contexts.

% Given the source code, we have implemented an automated script that can transform it into the S-TAC form by traversing its abstract syntax tree and performing the transformation recursively from the bottom up.

Taking the code shown in Listing~\ref{lst:gson17} and~\ref{lst:gson17donor} as examples, 
% the different operators (i.e., $==$ and $!=$) will not affects the matching of the two code
the S-TAC form for the if condition from the faulty code is $T_1:=in, peek()$, $T_2:=T_1, JsonToken.STRING$, while it will be $T_1:=in, peek()$, $T_2:=T_1, JsonToken.NULL$ for the
% condition from the 
reference code, where $T_1$ and $T_2$ are intermediate symbols. In this way, the two conditions can be better matched with each other. We will also evaluate the contribution of our S-TAC form in the experiment (Section~\ref{sec:result}). 

\subsection{Reference Code Adaptation}
\label{sec:adapt}

When given the faulty code, we can obtain a set of reference code based on the previous two processes. Then for each reference code pattern, \tool{} tries to generate candidate patches by matching the faulty code to the reference code. Algorithm~\ref{alg:matching} presents the details of the matching process. In general, \tool{} performs a greedy match between the faulty code and the reference code. Then, it generates patches according to the difference between them. In other words, given the token (or S-TAC) sequences for both faulty (i.e., \textit{BS}) and reference code (i.e., \textit{RS}), the matching algorithm will return a set of pairs (i.e., \textit{PS}), based on which \tool{} generates candidate patches according to a set of predefined rules.

\begin{figure}[t]
	\centering
	\includegraphics[width=0.98\columnwidth]{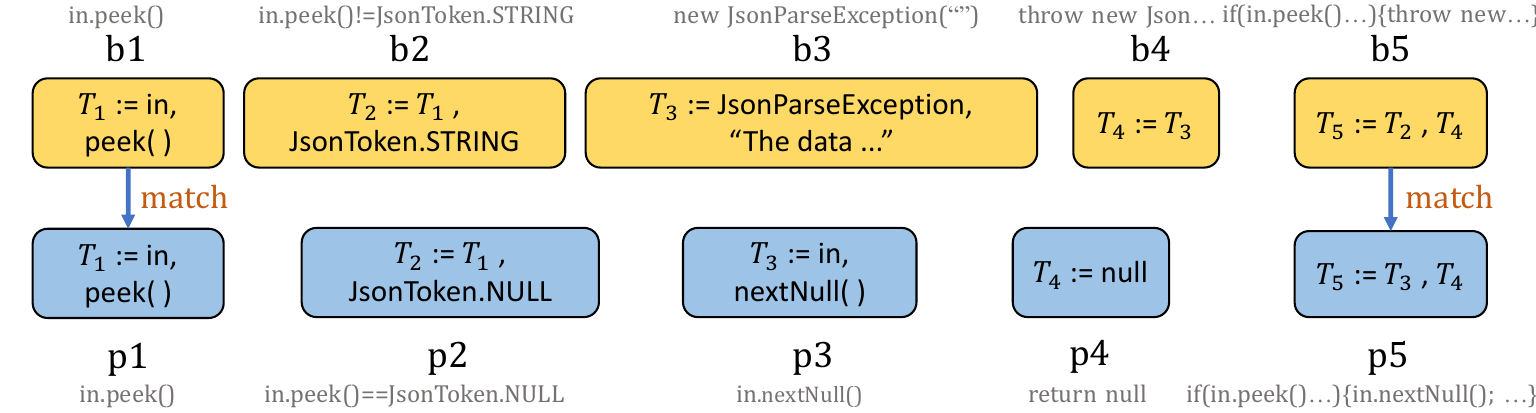}
        % \Description{Matching results of partial S-TACs in Gson-17 shown in Listing~\ref{lst:gson17}. $b_i$ represents the S-TAC form of the faulty \codeIn{if} statement while the $p_i$ corresponds to the \codeIn{if} statement in the reference code.}
	\caption{Matching results of partial S-TACs in Gson-17 shown in Listing~\ref{lst:gson17}. $b_i$ represents the S-TAC form of the faulty \codeIn{if} statement while the $p_i$ corresponds to the \codeIn{if} statement in the reference code.
 In particular, we also present the  source code corresponding to each T-SAC.}
	\label{fig:sat-mapping}
    % \vspace{-20pt}
\end{figure}

Given the token (or S-TAC) sequences of both faulty (\textit{BS}) and reference code (\textit{RS}), \tool{} first computes their longest common sequence (\textit{cs}, line 3) for greedy matching. It then establishes mappings between unmatched elements in the faulty and reference code based on these results (lines 5-13). Using \textit{S.getNodes(a, b)}, \tool{} retrieves elements between positions $a$ and $b$ in sequence $S$. As illustrated in Figure~\ref{fig:sat-mapping}, \textit{cs} for the S-TACs in Listings~\ref{lst:gson17} and~\ref{lst:gson17donor} is $[(b1, p1), (b5, p5)]$. From the figure we can observe that the S-TAC representation abstracts away concrete code content, mitigating its impact on pattern matching. The unmatched elements are divided into two lists, \textit{os} ($[b2, b3, b4]$) and \textit{ts} ($[p2, p3, p4]$), whose Cartesian product (i.e., $\Join$) forms all possible unmatched pairs (line 8). In particular, if no sibling AST nodes of a faulty code element are matched (line 21), \tool{} attempts to map its parent nodes via \textit{tryToMatchParent(*)} (lines 21-25). For instance, since the \codeIn{throw} statement ($b4$) has no matched siblings, \tool{} maps the faulty \codeIn{if} statement to its reference counterpart.

\begin{algorithm}[h]
	\footnotesize
	\caption{Reference Code Matching}
	\label{alg:matching}
	\SetKwInput{KwInput}{Input}                % Set the Input
	\SetKwInput{KwOutput}{Output}              % set the Output
	\DontPrintSemicolon
	
	\KwInput{\textit{BS:} \textcolor{gray}{a sequence of tokens or S-TACs that represent the bug.} 
	\textit{RS:} \textcolor{gray}{a sequence of tokens or S-TACs that represent the reference code.}}
	\KwOutput{\textit{PS:} \textcolor{gray}{a set of original and target node pairs.}}

	% Set Function Names
	\SetKwFunction{FMain}{matchElement}
        \SetKwFunction{FBuild}{tryToMatchParent}
	
	\SetKwProg{Fn}{Function}{:}{}
	\Fn{\FMain{\textit{BS, RS}}}{
		\textit{PS} $\leftarrow  []$, \textit{bf} $\leftarrow  -1$, \textit{rf} $\leftarrow  -1$  \;
            \textit{cs} $\leftarrow LCS(BS,RS)$  \tcp{The longest common sequence}
            \uIf{ cs.size() $>$ 0}{
                \ForEach{c in cs}{
                    \textit{os} $\leftarrow BS.getNodes(bf,BS.indexOf(c.fst))$ \;
                    \textit{ts} $\leftarrow RS.getNodes(rf,RS.indexOf(c.snd))$ \;
                    \textit{PS.append(os$\Join$ts)} \tcp{All unmatched pairs}
                    \textit{bf} $\leftarrow BS.indexOf(c.fst) $ \;
                    \textit{rf} $\leftarrow RS.indexOf(c.snd) $ \;  
                }     
                % \textit{tail} $\leftarrow cs.getLastNode()$ \;
    	    \textit{os} $\leftarrow BS.getNodes(bf,BS.length)$ \;
                \textit{ts} $\leftarrow RS.getNodes(rf,RS.length)$ \;
                \textit{PS.append}(os$\Join$ts) \tcp{Unmatched pairs at the end}
            }
            \textit{PS $\leftarrow$ PS $\cup$ \FBuild{PS}}\;
	    \KwRet \textit{PS}
	}	
        \Fn{\FBuild{\textit{PS}}}{
            \textit{result} $\leftarrow  []$ \;
            \ForEach{$\left \langle a,b \right \rangle $ in PS}{
                \textit{parent} $\leftarrow a.getParent()$ \;
                \textit{children} $\leftarrow parent.getChildren()$ \;
                \uIf{PS.FirstSet().containsAll(children)}{ \tcp{\add{All sibling AST nodes are not matched}}
                    \textit{ts} $\leftarrow  []$ \;
                    \ForEach{c in children}{
                    \textit{ts.append(PS.getSecond(c).getParent())}\;
                    }
                    \textit{result.append}([parent] $\Join$ ts)\; \tcp{\add{Add mappings of parent AST nodes}}
                }
            }
		\KwRet \textit{result}
	}
\end{algorithm}
% \vspace{-10pt}

% For the elements in the reference, \tool{} creates a mapping with the bug block. The key is each element and the value is its index in the bug sequence if it matches, or -1 if it does not match. 

Finally, according to the constructed mapping (i.e., \textit{PS}), \tool{} will generate candidate patches from each pair $(a,b)\in PS$ according to the following rules (we use $a$ to represent the associated AST node in what follows for ease of presentation): (1) replace $a$ with $b$ if the value type of $b$ is compatible with that of $a$; (2) insert $b$ before and after $a$ if $b$ is a standalone statement, e.g, expression statement; (3) insert $b$ before $a$ if $b$ is a conditional expression.
Both token-level and expression-level patch generations follow this mapping algorithm, based on the types of $a$ and $b$ (e.g., replacing variable $a$ with another variable $b$ generates a token-level patch).
% \jcom{to check and refine.} 
In particular, we disable delete operations for patch generation since they tend to generate incorrect patches~\cite{tananti,ISSTA18-SimFix}. Moreover, during this process, \tool{} \cradd{incorporates a built-in validation step that performs lightweight static analysis to check the validity of the newly generated code $b$}\crdel{will automatically check the validity of the new code $b$ via static analysis for efficiency}, such as whether all variables used by $b$ are valid and usable in the faulty location by checking their scopes.
\subsection{Patch Ranking}
\label{sec:rank}
% \z{Building the prefix trees takes quite some computational costs; why not utilize the frequencies collected in the trees for the expression-level candidate ranking too? How are the patch candidates returned by the token-level mining and the expression-level search ranked among each other? Although Section 4.3 says "all the patches will be ranked based on the ranking strategy introduced in Section 3.4.", however, Section 3.4 gives two different ranking scores for the token-level mining and the expression-level search, but doesn't say how the two different kinds of ranking scores are compared.}
After generating candidate patches, they will be evaluated against the test suites associated with the faulty project. To make the correct patches be evaluated as early as possible and reduce incorrect patches, we have proposed a hierarchical patch ranking strategy. As demonstrated in our preliminary study, most of the reusable code elements exist at the fine-grained granularity. Furthermore, previous study \cite{DirectFix} also shows that correct patches tend to involve small code changes. Therefore, \tool{} ranks all the patches generated by the finer-grained token-level code changes higher than those generated by the coarse-grained expression-level code changes.
% In other words, the patches generated by token-level code changes will be evaluated first and then those by expression-level code changes. Please note that the online pattern mining process introduced in Section~\ref{subsec:mining} can be proceeded independently ahead of time and the constructed patterns can be reused for repairing different bugs on demand.

% We incorporate this prior knowledge by measuring the complexity of candidate patches. 
%Specifically, \tool{} ranks all the patches generated by the coarse-grained expression-level code changes higher than those generated by the token-level code changes. The reasons are twofold: (1) The number of reference codes for expression-level program repair is much smaller than that for token-level repair.\jun{Conflict with the fine-grained mining.} (2) Intuitively, the coarse-grained code changes tend to have larger effects on the program semantics, and thus the incorrect patches can be more easily identified by the test suite for improving the patch precision.

Then, regarding the patches generated by expression-level code changes, \tool{} simply takes the similarity in code search (refer to Section~\ref{sec:online}) as the ranking score by following previous studies~\cite{ISSTA18-SimFix}. For patches generated by token-level code changes,
% \add{Based on prior works~\cite{DirectFix,ISSTA18-SimFix} and the motivation derived in Section~\ref{sec:motivation}, more frequent patterns have higher possibility to produce correct patches and most of patches tend to require finer-grained modifications.}
% for patches generated by token-level code changes, we aim to rank patches that utilize more frequently occurring patterns and require finer-grained modifications higher.
 we consider two important factors: the frequency of the reference code pattern and the similarity between the faulty and fixed code by following existing studies~\cite{DirectFix,ISSTA18-SimFix}.
%Specifically, 
%the patches generated by more frequent patterns will rank higher, because frequent patterns may have a larger possibility to reflect the programming conventions. Therefore, we will be more confident to perform the code change according to the patterns. On the other hand, as reported by previous studies~\cite{DirectFix}, the correct patches tend to involve small code changes. Therefore, we also incorporate such insight for patch ranking. Finally, 
The ranking score is defined by Formula~\ref{eq:score},%, which evenly weighs the importance of these two factors. 
where $freq$ denotes the frequency of a pattern and $max\_freq$ denotes the max $freq$ of patterns referenced for generating patches. $L_{orig}$ and $L_{fixed}$ represent the length of source code (i.e., number of tokens) before and after applying a patch, while \textit{LevenshteinDist} represents the token-level Levenshtein distance~\cite{levenshtein1966binary} between the faulty and the fixed code.
As a result, a patch with a higher score (i.e., referencing a more frequent pattern and applying finer-grained modifications) will rank higher and thus will be validated earlier by running the test cases in the faulty project.
\begin{equation}\label{eq:score}
\small
 score = 0.5 * \frac{freq}{max\_freq} + 0.5 * (1 - \frac{{LevenshteinDist}}{max({L_{orig}}, {L_{fixed}})})
\end{equation}

% \begin{equation}\label{eq:score}
%  score = 0.5 * frequency + 0.5 * similarity
% \end{equation}
% \begin{equation}
%    frequency = \frac{occurrence}{max\_occurrence}
% \end{equation}
% \begin{equation}\label{eq:dis}
%    similarity = 1 - \frac{{LevenshteinDist}}{max({L_{original}}, {L_{fixed}})}
% \end{equation}

% $$score = 0.5*\rm pattern score + 0.5 * editscore$$
% $$patternsocre = \frac{occurrence}{max\_occurrence}$$
% $$editscore = 1 - \frac{\rm{LevenshteinDistance}}{Max\{\rm{originalCodeLength}, \rm{fixedCodeLength}\}}$$

\section{Experiment Configuration}
\label{sec:exp}

% \subsection{Research Questions}
% \label{sec:rq}
We address the following research questions in the study.

\begin{itemize}[leftmargin=*]
	\item \textbf{RQ1:} How effective is \tool{} in repairing real-world bugs \cradd{, compared to the state-of-the-art APRs}?

	\item \textbf{RQ2:} Can \crdel{our approach}\cradd{a combination of Repatt with traditional APR approaches} improve the state-of-the-art APRs?
	
	\item \textbf{RQ3:} What is each component’s contribution in \tool{}?
 % to the effectiveness of our approach?

	\item \textbf{RQ4:} What is the impact of the pattern frequency in \tool{}?

%        \z{
%        \item \textbf{RQ4:} How effective when combining our method with existing redundancy-based APR technologies?
%
%        \item \textbf{RQ5:} How effective is our patch ranking strategy?}
%	% \item \textbf{RQ4:} What is the impact of the skip strategy in the code base construction?
%	
%	% \item \textbf{RQ5:} How to prove the validity of S-TAC?
%        	
\end{itemize}

%\hy{can merge RQ5 with RQ2? }

\subsection{Subjects and Baselines}
\label{sec:subject}

% \z{Newest techniques such as KNOD and ChatRepair are not considered.}
In our evaluation, we employed the widely-used Defects4J benchmark~\cite{just2014defects4j}, following existing studies~\cite{ssFix,xuan2016nopol,xiong-icse17,Yang2023transplantFix,jiang2023impact,zhu2021syntax,ISSTA18-SimFix,liu2019tbar}. 
% The details of the benchmark is presented in Table~\ref{tab:subject}. 
In particular, we used both Defects4J v1.2 and v2.0 to demonstrate the generality of our approach. Specifically, v1.2 includes 395 bugs from six large-scale real-world projects and v2.0 includes 440 additional bugs from 12 real-world projects.

%all the bugs in the latest version v2.0 of this benchmark, which in total consists of 835 real-world bugs from 17 large-scale Java projects. \jun{Only one subject.}

% \input{table/subject}

Furthermore, to show the effectiveness of \tool{}, 
we compared it with \crdel{nine}\cradd{ten} state-of-the-art APR approaches from different categories, i.e., \cradd{CapGen~\cite{Wen2018ContextAwarePG},} SimFix~\cite{ISSTA18-SimFix}, TransplantFix~\cite{Yang2023transplantFix}, TBar~\cite{liu2019tbar}, Recoder~\cite{zhu2021syntax}, SelfAPR~\cite{ye2022selfapr}, ITER~\cite{ye2024iter}, AlphaRepair~\cite{xia2022less}, Repilot~\cite{wei2023copiloting} and GAMMA~\cite{zhang2023gamma}. 
Specifically, \cradd{CapGen, }SimFix and TransplantFix are the \crdel{two}\cradd{three} latest and representative \textbf{redundancy-based} methods that repair bugs by referencing similar code. TBar is the best-performing \textbf{template-based} repair technique that generates candidate patches by a set of manually defined patch templates. Recoder, ITER and SelfAPR are the latest and best-performing \textbf{deep learning methods} that are specially designed for program repair. Finally, AlphaRepair, Repilot, and GAMMA represent the most recent advance in APR techniques by \textbf{adopting LLMs}. These baselines are the best-performing APR approaches using different technologies and their complete experimental results are available. By comparing our approach with these diverse baselines, we would like to analyze the overall effectiveness of our approach from different perspectives and make the conclusions reliable.

%For the experiment, in order to prove our generalizability, we evaluate our method on both Defects4J v1.2 and Defects4J v2.0 database, including 835 reproducible bugs from real-world open-source projects.

\subsection{Configuration and Metrics}
\label{sec:conf}
For each bug, \tool{} first constructs the token-level code patterns based on the complete faulty program under repair before the online repair (see Section~\ref{subsec:mining}). The construction is actually efficient and on average took about three minutes in our experiment. Then, given the faulty line of code, \tool{} first generates at most 200 patches based on the constructed token-level code patterns and then generates at most 1000 patches based on the expression-level code patterns due to its large search space. Finally, all the patches will be ranked based on the ranking strategy introduced in Section~\ref{sec:rank}.
% the max number of reference snippets from online code search as \jcom{2000}, and 
%the max number of candidate patches as {1000} for expression-level repair and {200} for token-level repair for each location.
% Actually, almost all bugs can be correctly repaired within 90 minutes, which assigns with existing studies~\cite{Xuan2016History,Wen2018ContextAwarePG,ELIXIR}
% Additionally, we assign each bug 90 minutes\jun{Incorrect, we set the number of maximum patches for evaluation as 1000. In fact, all the bugs can be repaired within 90 minutes.} for repair~\cite{Xuan2016History,Wen2018ContextAwarePG,ELIXIR}. We empirically set the default value of {MIN\_SUPPORT} in token-level pattern mining as 3, and the MAX\_SKIP and MAX\_LEN as {8} and {5}, respectively. 
For baselines, we adopt their published experimental results directly in their open-source repositories. \cradd{Specifically, as the original CapGen does not provide experimental results on the Closure and Mockito bugs from Defects4J~\cite{just2014defects4j}, we supplement these results using those results reported in a previous study~\cite{ghanbari2018practical}}

In our experiment, \tool{} generates at most 3 candidate patches that can pass all the test cases (i.e., plausible patches) for each bug since some baseline approaches did not report the ranking of patches, e.g., SelfAPR and LLM-based methods. Following previous study~\cite{ssFix,ISSTA18-SimFix,liu2019tbar,zhu2021syntax}, a patch is deemed to be correct \textit{iff} it is semantically equivalent to the developer patch by manual check. \cradd{In this process, the first three authors independently conducted the annotation and reached a consensus through discussion.} We also published all our results for further inspection and verification. 
% \add{In particular, authors labeled the plausible patches independently and we also used the Cohen's Kappa coefficient~\cite{cohens} to measure the inter-rater agreement between them. The result is always 1.0, i.e., perfect agreement.}
% we manually check the correctness of candidate patches. 
% A patch is considered as correct only if it is semantically equivalent to the developer patch. 
%Following existing studies~\cite{ssFix,ISSTA18-SimFix,liu2019tbar,zhu2021syntax}, 
Finally, we report the number of correctly repaired bugs (equivalent to the well-known \textbf{recall}) and the \textbf{precision} of patches (the ratio of correctly repaired bugs to all the bugs that have plausible patches).
% Please note the \textbf{recall} and \textbf{precision} here are different from those in the deep learning field since all the bugs are expected to be repaired, i.e., the \textit{true negative} in deep learning field does not make sense here.

%\z{In the experiment for RQ4, to demonstrate how our patch ranking strategy can help to choose the correct patches, we only consider the top-ranking patches.
%We deem a patch as a correct one when (1) it can pass all human-written testsuites and (2) it is identical to the developer patch or if it is considered as correct by manual analysis.
%}
All experiments were conducted on a server with Ubuntu 18.04, equipped with 128GB RAM and a processor of Intel(R) Xeon(R) E5-2640.

\section{Result Analysis}
\label{sec:result}
%\jun{
%\begin{itemize}
%    \item How many patches per each bug? Compilation rate and time cost.
%    \item More analysis, more insights, examples.
%    \item Combining existing methods due to complementary.[done]
%    \item Not high precision in RQ1.
%\end{itemize}
%}

\begin{table*}[!t]
\caption{Number of bugs repaired by different methods with perfect fault localization. In the table, x/y denotes the corresponding approach generates correct patches for x bugs and generates plausible patches for y bugs.
%	\hy{The results in Table 1 show that RePatt may not be the best?}
%	\jcom{@Zhirui, add the results of learning-based methods. If no enough space, we can report the overall results on D4J1.2 and D4J2.0. respectively.}
}	\label{tab:comparisonPerfect}
	\centering
	% \small
	% \setlength\tabcolsep{3pt}
	\resizebox{.98\textwidth}{!}
   {

\begin{tabular}{cl|c|c|c|c|c|c|c|c|c|c|c}
\toprule
&  {\bf Project}  &{\bf  TBar}    &{\bf SimFix}  & {\bf TransplantFix}   & {\bf \del{SelfRepair}{SelfAPR}}  & {\bf  Recoder}     & {\bf AlphaRepair} & {\bf Repilot} & {\bf GAMMA} &  {\cellcolor{lightgray}{\bf \tool{}}} &  {\cellcolor{lightgray}\bf{ \combine{}}}\\
 \midrule
 \multirow{6}{*}{\makecell{Defects4J v1.2 \\ (395 bugs)}} &
  Mockito           & 3/3           & 0/0     & 3/3       & 3/3   & 2/-             & 4/-       & 0/-       & 2/-    & \cellcolor{lightgray}{\bf 1/2}       & \cellcolor{lightgray}\bf{ 2/5} \\
& Closure           & 16/24         & 5/6     & 10/19     & 19/23 & 23/-            &   22/-       & 21/-      & 23/-   & \cellcolor{lightgray}{\bf 10/12}     & \cellcolor{lightgray}\bf{20/31} \\
& Chart             & 11/13         & 4/8     & 6/10      & 7/10   & 10/-            & 8/-       & 5/-       & 10/-    & \cellcolor{lightgray}{\bf 6/7}       & \cellcolor{lightgray}\bf{ 11/21} \\
& Lang              & 13/18         & 8/13    & 4/10      & 10/13  & 10/-            & 12/-       & 14/-       & 15/-    & \cellcolor{lightgray}{\bf 6/11}      & \cellcolor{lightgray}\bf{ 16/27} \\
& Math              & 22/35         & 14/26   & 12/25     & 21/25  & 18/-            & 20/-     & 20/-     & 24/-  & \cellcolor{lightgray}{\bf 15/25}     & \cellcolor{lightgray}{\bf 22/51} \\
& Time              & 3/6           & 1/1     & 1/2       & 3/3   & 3/-             & 2/-       & 1/-       & 2/-    & \cellcolor{lightgray}{\bf 2/3}       & \cellcolor{lightgray}{\bf 4/7} \\ 
\midrule \multirow{12}{*}{\makecell{Defects4J v2.0\\ (440 bugs)}} 
& Closure           & 0/0           & 1/1     & 0/3       & 1/1     & 0/-             &0/-        & 0/-       &0/-     & \cellcolor{lightgray}{\bf 1/1}       & \cellcolor{lightgray}{\bf 2/5} \\
& Cli               & 1/7           & 0/1     & 4/8       & 8/9   & 3/-             & 5/-       & 6/-       & 9/-    & \cellcolor{lightgray}{\bf 4/9}       & \cellcolor{lightgray}{\bf 4/15} \\
& Codec             & 3/6           & 1/1     & 2/4       & 8/9   & 2/-             & 6/-       & 6/-       & 3/-    & \cellcolor{lightgray}{\bf 4/6}       & \cellcolor{lightgray}{\bf 3/8} \\
& Collections       & 0/0           & 0/0     & 0/0       & 1/1   & 0/-             & 0/-   &1/-    &   0/-   & \cellcolor{lightgray}{\bf 0/0}       & \cellcolor{lightgray}{\bf 0/0} \\
& Compress          & 2/12          & 0/4     & 4/10      & 6/8   & 3/-             &   1/-    &   3/-    &   4/-        & \cellcolor{lightgray}{\bf 4/13}      & \cellcolor{lightgray}{\bf 7/18} \\
& Csv               & 2/6           & 0/2     & 1/2       & 1/1   & 3/-             & 1/-    &   3/-    &   0/-    & \cellcolor{lightgray}{\bf 1/3}       & \cellcolor{lightgray}{\bf 2/7} \\
& Gson              & 1/4           & 2/2     & 1/2       & 1/1   & 0/-             & 2/-    &   1/-    &   3/-     & \cellcolor{lightgray}{\bf 2/3}       & \cellcolor{lightgray}{\bf 2/5} \\
& JacksonCore       & 0/5           & 0/2     & 1/6       & 3/3   & 0/-             & 3/-    &   3/-    &   3/-    & \cellcolor{lightgray}{\bf 3/7}       & \cellcolor{lightgray}{\bf 3/9} \\
& JacksonDatabind   & 2/17          & 1/10    & 7/20      & 8/10  & 0/-              & 8/-    &   8/-    &   10/-    & \cellcolor{lightgray}{\bf 10/18}     & \cellcolor{lightgray}{\bf 11/33} \\
& JacksonXml        & 0/1           & 0/0     & 0/0       & 1/1   & 0/-               & 0/-  &   0/-  &   0/-    & \cellcolor{lightgray}{\bf 0/0}       & \cellcolor{lightgray}{\bf 1/1} \\
& Jsoup             & 6/17          & 1/2     & 3/8       & 6/8  & 7/-              & 9/-    &   18/-    &   11/-    & \cellcolor{lightgray}{\bf 5/13}      & \cellcolor{lightgray}{\bf 10/25} \\
& JxPath            & 1/6           & 0/0     & 2/7       & 1/1   & 0/-             & 1/-    &   1/-    &   2/-    &  \cellcolor{lightgray}{\bf 1/5}      & \cellcolor{lightgray}{\bf 4/12} \\ 
\midrule
\multicolumn{2}{c|}{Total}       
                    &  85/180        &  38/79  & 61/139    & 108/130  &84/-            & 104/-    & 111/-     & 121/- & \cellcolor{lightgray}{\bf 75/138 }   & \cellcolor{lightgray}{\bf 124/280}  \\ \midrule
\multicolumn{2}{c|}{Precision(\%) }          
                    &  47.2          & 48.1    & 43.9      & 83.1   & -              & -      & -      & -   & \cellcolor{lightgray}{\bf 54.3}      & \cellcolor{lightgray}{\bf 44.3}\\
\bottomrule
\end{tabular}
% \begin{tablenotes}
%     \item[*] SelfRepair did consider the ranking of patches.
% \end{tablenotes}	
  % \end{threeparttable}
}

\vspace{-10pt}
\end{table*} 

\subsection{Overall Effectiveness of \tool{} (RQ1)}
\label{sec:rq1}
\subsubsection{Perfect fault localization}

As explained in Section~\ref{sec:subject}, we evaluated the effectiveness of our approach by comparing it with eight state-of-the-art APR approaches since ITER was only evaluated under the automated fault localization setting (\textit{ref.} Section~\ref{sec:automated}). The repair results are presented in Table~\ref{tab:comparisonPerfect} when given the actual faulty locations by following existing studies~\cite{jiang2023impact,Yang2023transplantFix,zhu2021syntax,liu2019tbar,ye2022selfapr}.
% In each cell, \textit{X/Y} denotes the repair method can generate at least one plausible patch for \textit{Y} bugs on different projects, where \textit{X} bugs are indeed repaired correctly. 
From the table we can see that \tool{} successfully repaired 75 bugs while generating incorrect patches for other 63 bugs, yielding a patch precision of 54.3\%. \textbf{Notably, {66} bugs were correctly repaired by the first plausible patch}, suggesting that developers could focus primarily on the first patch generated by \tool{} to reduce manual validation efforts. The results also show that our approach outperforms the state-of-the-art redundancy-based methods SimFix and TransplantFix by repairing 97.4\% and 23.0\% more bugs, indicating the effectiveness of our approach in reusing code at different granularities for patch generation. Compared to SelfAPR, which repaired 108 bugs when considering Top-50 generated patches but only 34 when restricted to Top-1, \tool{} shows a more favorable trade-off between precision and recall. \cradd{When analyzing the incorrect patches generated by \tool{}, we found that most of them resulted from the identification of non-representative patterns, which led to incorrect modification locations or inappropriate fixes. However, due to the weakness of the test suites~\cite{xiong-icse18, tian2022predicting, PatchPlausibility}, these patches were able to pass all test cases, highlighting a common challenge in test-based APR evaluation.}

Similarly, LLM-based repair methods also did not report the ranking of their correct patches. Specifically, AlphaRepair~\cite{xia2022less} and Repilot~\cite{wei2023copiloting} generated up to 5,000 patches per bug, while GAMMA~\cite{zhang2023gamma} even did not limit the number of candidate patches.
% \add{Additionally, recent LLM-based repair tools impose loose constraints on patch generation(e.g., AlphaRepair~\cite{xia2022less} and Repilot~\cite{wei2023copiloting} generate up to 5,000 patches per bug, while GAMMA~\cite{zhang2023gamma} has no hard limit). 
Although these methods may achieve a higher number of correct fixes, they require developers to spend significant time manually reviewing patches, reducing their practicality~\cite{PatchPlausibility,xiong-icse18}. In contrast, \tool{} enforces strict limits on the number of generated patches, effectively mitigating this issue.
Furthermore,
% to evaluate the effectiveness of \tool{}’s expression-level search, 
we analyzed \tool{}'s performance on repairing multi-location bugs. Among 75 correctly repaired bugs, 30 involved multiple locations, highlighting the effectiveness of \tool{}’s fine-grained repair strategy.
% Additionally, the latest learning-based repair tool KNOD~\cite{jiang2023knod} did not report detailed Top-1 patch results. However, even when considering Top-10 patches, KNOD repaired only 20 and 13 bugs from Defects4J v1.2 and v2.0, respectively, underperforming compared to our approach. Compared with the others, our approach can also achieve similar effectiveness. For example, \tool{} achieved higher patch precision than TBar although repaired fewer bugs. In contrast, \tool{} repaired more bugs than the pre-trained models (CodeT5, CodeGen, and PLBART) but achieved a lower patch precision. 
% \add{Furthermore, to demonstrate the effectiveness of expression-level search in \tool{}, we conducted a further analysis of \tool{}'s performance on multi-location bugs. The results show that out of 75 correctly repaired bugs, 30 are multi-location bugs. This highlights the effectiveness of expression-level repair code search used in \tool{}.}

% delete for icsme
% Nevertheless, although our approach can effectively repair many bugs, there are still many bugs that cannot be repaired by \tool{} because the patches cannot be assembled from individual reference code. To further improve the correct repair, enlarging the patch space may help, such as combining multiple reference code or searching a code base beyond the faulty project for patch generation. We plan to further investigate its performance in the near future. 
%\hy{see if there are other metrics that show the benefits of RePatt? like the top-1 correct patch, etc}

\begin{figure}[h]
    \vspace{-10pt}
	\centering
    \subfigure[Traditional methods]{ \includegraphics[width=0.4\columnwidth]{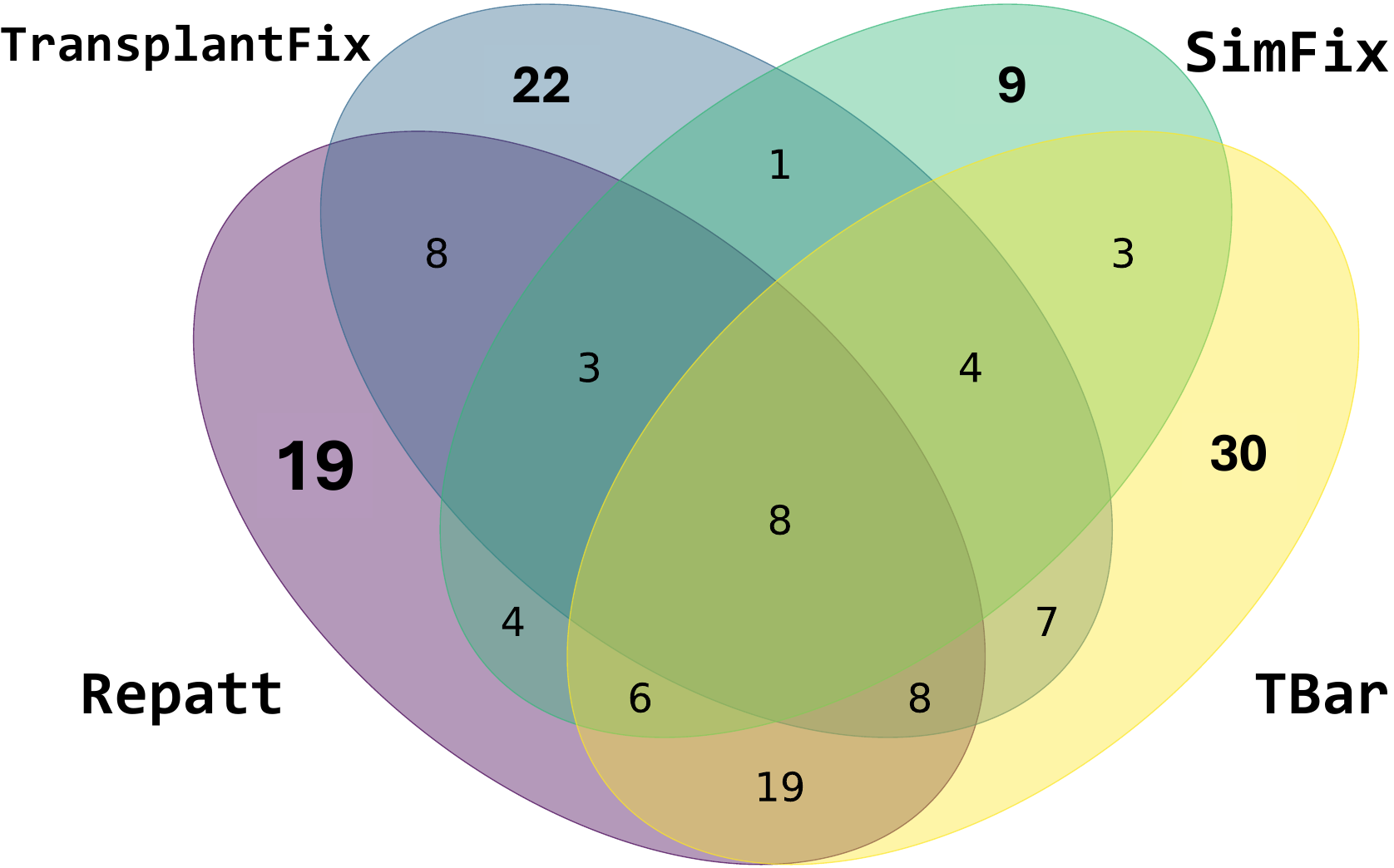}
    }
    \subfigure[LLM and DL based methods]{\includegraphics[width=0.42\columnwidth]{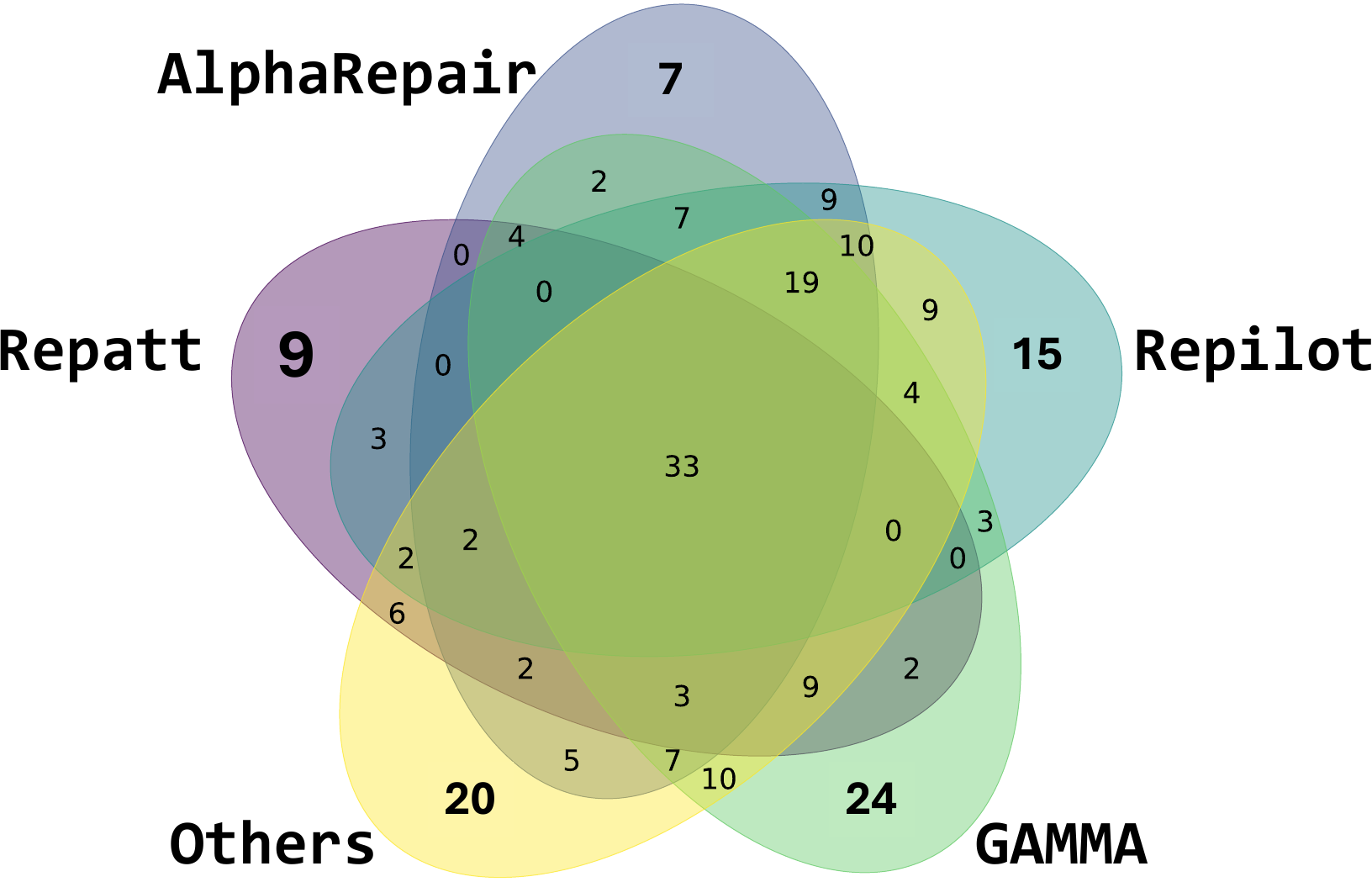}
    }
    % \hspace{-5pt}
    \subfigure[All methods]{\includegraphics[width=0.5\columnwidth]{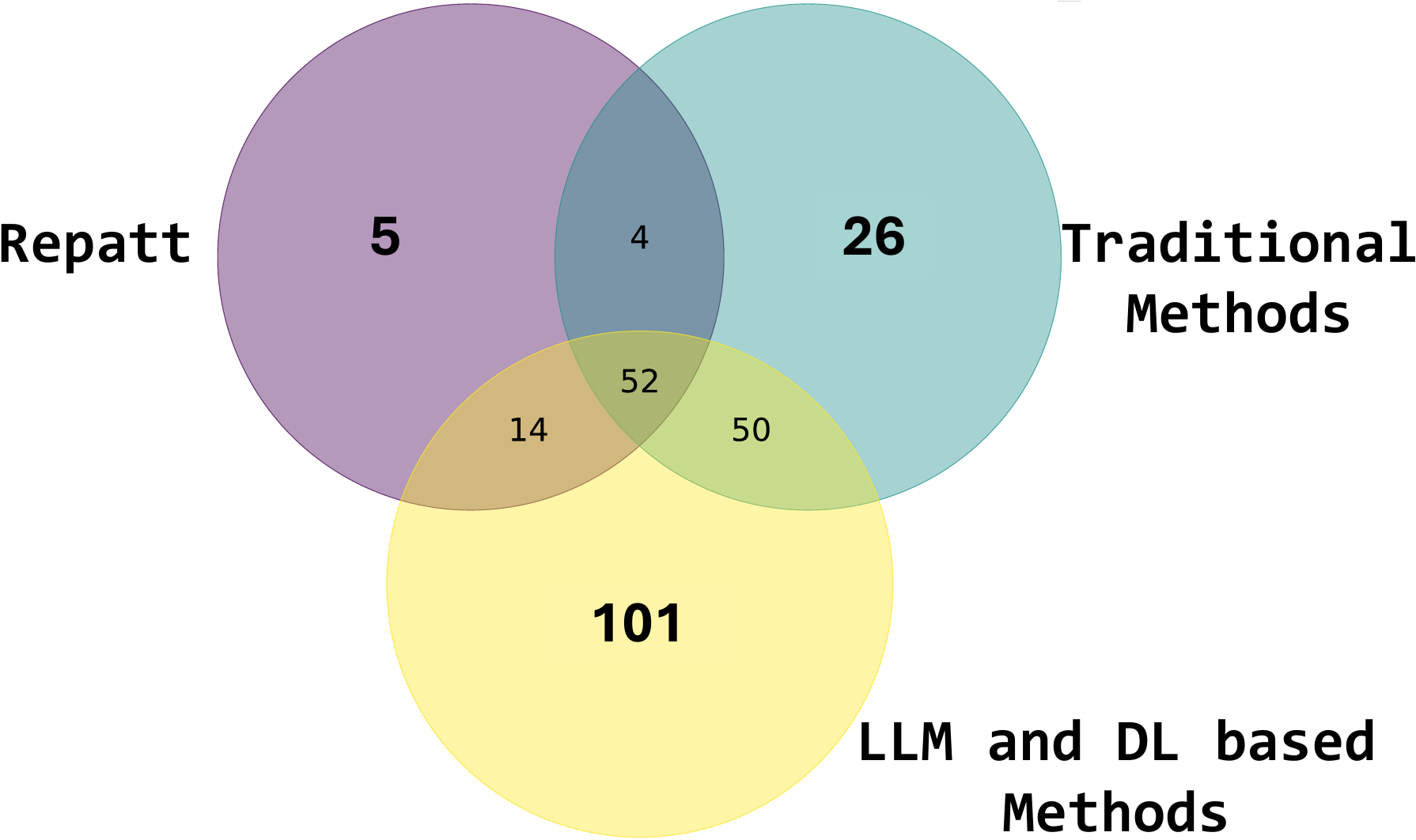}
    }
 % \includegraphics[width=0.99\columnwidth]{fig/overlap.png}
        % \Description{Overlaps of bugs repaired by different approaches.}
	\caption{Overlaps of bugs repaired by different approaches.}
	\label{fig:overlap}
    % \vspace{-10pt}
\end{figure}

\subsubsection{Degree of complementary}\label{sec:comp}
% According to the results shown in Table~\ref{tab:comparisonPerfect}, although \tool{} can successfully repair many bugs, it is unknown whether \tool{} is still valuable in the face of many state-of-the-art APR approaches. Therefore, we further investigated the extent to which our approach complements the state-of-the-art baselines. 
We also analyzed the overlaps of bugs repaired by both \tool{} and the baselines. Specifically, we classified the baselines into traditional, LLM-based and deep-learning-based methods according to the patch generation techniques, and then compared \tool{} with these two methods separately. Figure~\ref{fig:overlap} presents the results. In summary, \tool{} successfully repaired {19} unique bugs that the traditional APRs cannot fix, and {9} unique bugs that the LLM-basd and deep learning-based methods cannot fix. In particular, when compared with all the eight baselines, our approach still can repair {5} unique bugs.
% \del{For example, to repair the bug shown in Listing~\ref{lst:codec3}, the constant value ``4'' should be replaced by ``3''. Without the guidance of the fine-grained reference code, it is nearly impossible for existing APRs to get the correct patch due to the large search space. In general, compared with traditional APRs, \tool{} can more effectively utilize fine-grained reusable code elements for patch generation, while compared with deep learning methods, \tool{} only depends on a small number of reusable samples rather than the large-scale training data, which enables \tool{} to repair infrequent bugs. In conclusion, the results reflect that our approach indeed complements existing approaches.}
For example, to repair the bug shown in Listing~\ref{lst:codec3}, the constant value ``4'' should be replaced by ``3''. Without the guidance of the fine-grained reference code, it is nearly impossible for existing APRs to get the correct patch due to the large search space. In general, compared with traditional APRs, \tool{} can more effectively utilize fine-grained reusable code elements for patch generation\crdel{,}\cradd{. To better understand its repair capabilities, we manually analyzed the 19 bugs that were uniquely fixed by \tool{}. We found that 7 of them (36.8\%) involved fine-grained code reuse and edits, such as changing a single character, a variable name or a method name. This demonstrates the capability of \tool{} to precisely identify and apply fine-grained code changes effectively, which traditional approaches may overlook due to their coarse-grained search spaces. Moreover,} while compared with LLM-based and deep learning-based methods, \tool{} only depends on a small number of reusable samples rather than the large-scale training data, which enables \tool{} to repair infrequent bugs, especially those requiring domain-specific knowledge. In conclusion, the results reflect that our approach indeed complements existing approaches.

\subsubsection{Automated fault localization}
\label{sec:automated}
In addition, some of the baseline approaches~\cite{ISSTA18-SimFix,liu2019tbar,zhu2021syntax, Yang2023transplantFix, ye2024iter} were also evaluated in a more realistic scenario where the perfect fault localization was unknown. To ensure a fair comparison, we evaluated our approach under the same conditions. Specifically, we utilized the commonly-used Ochiai~\cite{abreu2007accuracy} algorithm, implemented by the GZoltar toolset~\cite{Campos2012GZoltar}, to obtain a list of candidate faulty locations like existing APR tools~\cite{liu2019tbar,ISSTA18-SimFix,zhu2021syntax}. Table~\ref{tab:sbfl_result} summarizes the experimental results. From the table, we can see that the number of correctly repaired bugs drops sharply for all the repair approaches
due to the inaccurate fault localization results.
% For example, SimFix, TBar, TransplantFix, Recoder, and \tool{} repaired 5.2\%, 41.2\%, 27.9\%, {20\%} and {60\%} fewer bugs, respectively, compared with providing the perfect fault localization. The large decline in our approach is mainly due to the inaccurate fault localization results, which will cause a large search space since the finer-grained code reuse tends to generate candidate patches, making the correct patches unfound within the given time budget. 
Moreover, almost all baseline approaches
% \add{(except ITER, which is not suitable for evaluation in the perfect fault localization scenario)} 
experienced a significant decline in patch precision. For instance, the precision of SimFix, TBar and TransplantFix drops {7.8\%}, {19.1\%}, and {26.9\%}, respectively. In contrast, the precision of our approach was improved by {54.3\%}.
% \add{Additionally, we studied the performance of \tool{} without applying the patch ranking component mentioned in Section~\ref{sec:rank}. As shown in the column $\tool{}_{w/o \ ranking}$ of Table~\ref{tab:sbfl_result}, \tool{} is still able to generate at least one plausible patch for 36 bugs, with 23 of these bugs being correctly fixed. This results in a precision of 63.9\%, which is still higher than the compared baselines.}
\add{One of the reasons is that fine-grained code changes focus on fixing local code without making extensive changes to the original content, thereby preserving program logic and maintaining the effectiveness of test cases, leading to more correct patches. In contrast, coarse-grained modifications often involve larger code changes, which may unintentionally alter program functionality, resulting in many patches that pass the tests but are semantically incorrect due to the problem of weak tests~\cite{PatchPlausibility,xiong-icse18,xin2017identifying}.}
% One of the reasons is that fine-grained code changes at the faulty location directly affect the results of failing tests and thus have a higher possibility to make them pass with lower possibility affecting the results of passing tests, while fine-grained code changes to the correct code will have a larger impact on the passing tests. In contrast, the coarse-grained code changes by existing APRs tend to affect both failed and passed tests, and thus they will generate incorrect patches more easily under inaccurate fault localization due to the problem of weak tests~\cite{PatchPlausibility,xiong-icse18,xin2017identifying}.
In summary, our approach significantly outperforms the baselines by achieving 15.6\%-51.7\% higher patch precision.
Notice that the high precision of patches is very important since it affects the usability of APR techniques~\cite{PatchPlausibility,xiong-icse18}, especially in the real-world repair scenarios where perfect fault localization is unknown -- high precision denotes less wasted human effort for manual validation.

\begin{table}[!t]
\caption{Repair result when using SBFL (\#correct/\#plausible).}
% In the table, x/y denotes the corresponding approach generates correct patches for x bugs and generates plausible patches for y bugs.}
	\label{tab:sbfl_result}
\centering
\resizebox{\columnwidth}{!}{
\setlength\tabcolsep{3pt}
\begin{tabular}{l|c|c|c|c|c|c|c|c}
\toprule
\bf{} & \bf{\cradd{CapGen}} & \bf{SimFix} & \bf{TBar} & \bf{TransplantFix} & \bf{Recoder} & \bf{ITER}& \cellcolor{lightgray}\bf{\tool{}}&\cellcolor{lightgray}\bf{\combine{}}\\ 
\midrule
Total     & \cradd{22/49} & 36/81           & 50/131        & 44/137                 & 70/140    & 45/66           & \cellcolor{lightgray}\bf{31/37} & \cellcolor{lightgray}\bf{90/236}          \\
Precision (\%)  & \cradd{44.90}    & 44.4            & 38.2          & 32.1                   & 50.0           & 68.2             & \cellcolor{lightgray}\bf{ 83.8} &  \cellcolor{lightgray}\bf{38.1}   \\ \bottomrule      
\end{tabular}
% \end{threeparttable}
}
\end{table}

\subsection{Improvement over SOTA APRs (RQ2)}

As presented in Section~\ref{sec:comp}, \tool{} complements existing APR approaches. In this research question, we aim to explore the possibility of our approach to improving existing APRs. Specifically, we combine \tool{} with existing APRs by proposing a post patch ranking strategy and see whether it can further improve the best-performing APR. 
Given the patches generated by different APRs for a bug, we will rank the ones with smaller code changes higher by following the insights from both our study and the previous research~\cite{DirectFix}. In particular, we use GumTree~\cite{GumTree}, a fine-grained and mature code differencing tool, to measure the change sizes of patches. If two patches have the same number of code changes, we use the patch precision to break the tie. In this study, we combined our approach with the three traditional APR approaches, i.e., Simfix, TBar, and TransplantFix. In this way, the combined method shares a similar repair pipeline to the individuals, i.e., not demanding large-scale training data. Therefore, the patch ranking will be \tool{}$>$SimFix$>$TBar$>$TransplantFix based on their patch precision in Tables~\ref{tab:comparisonPerfect} and \ref{tab:sbfl_result} for patches having the same change size.

\begin{figure}[t]
	\begin{center}
	\subfigure[Using perfect FL]{\includegraphics[width=0.35\columnwidth]{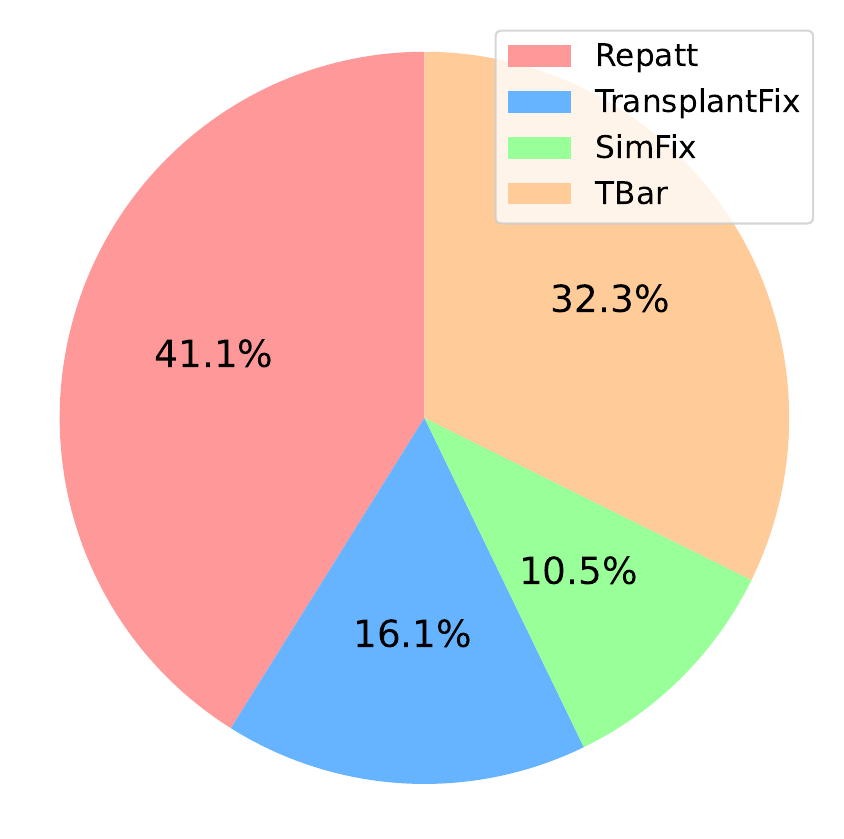}}\hspace{20pt}
	\subfigure[Using SBFL]{\includegraphics[width=0.35\columnwidth]{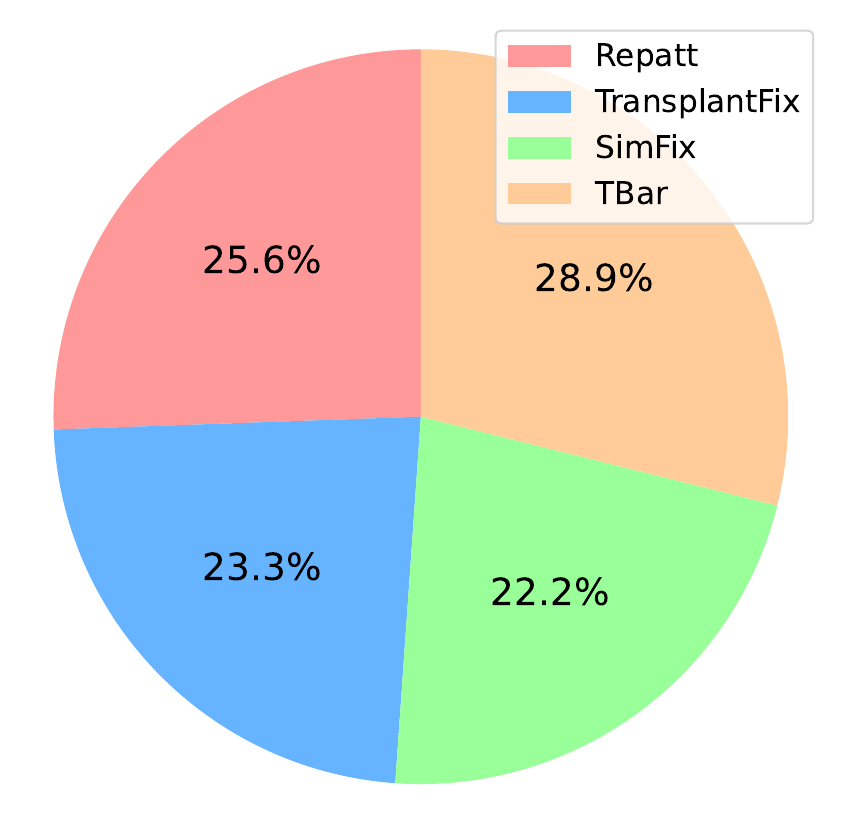}}
	\end{center}
        % \Description{Source of correct patches in \combine{}}
	\caption{Source of correct patches in \combine{}}
 \label{fig:patch_source}
 \vspace{-10pt}
\end{figure}

The experimental results are also presented in Tables~\ref{tab:comparisonPerfect} and~\ref{tab:sbfl_result} (i.e., \textbf{\combine{}}) when using the perfect fault localization and the SBFL, respectively. The results show that by combining the strength of different APR tools, \combine{} correctly repaired 124 bugs by the \textbf{first patches}, 
39 more bugs compared with the best-performing APR (TBar for Top-1 patches), leaving the improvement as 45.9\%. Moreover, even compared to LLM-based methods, which do not report detailed patch rankings and impose loose constraints on the number of generated patches (e.g., producing 5,000 or more patches per bug), \combine{} still outperforms these 3 LLM-based approaches, repairing up to 20 more bugs. Similarly, when using the SBFL results, \combine{} also outperforms the best-performing Recoder by repairing 20 more bugs, and improves the individuals by at least 80\% (\textit{vs} Tbar). Our results show the promise to improve the repair capability of APRs by combining the strength of individuals. 
\cradd{Although this approach may increase the time required to repair one defect, the additional overhead can be mitigated through parallel execution, similar to some LLM-based APR methods~\cite{xia2023plastic}.}
In particular, to analyze the contribution of our approach in the combination, we present the percentages of correct patches generated by each APR in Figure~\ref{fig:patch_source}. 
% From the figure we can see that 
About 41.1\% and 25.6\% correct patches were contributed by \tool{} under different repair settings, indicating its large contribution to the overall effectiveness of the combined method. However, though the combined method is effective in repairing much more bugs, the patch precision is still low, i.e., less than 45\%. To improve the patch precision, existing patch filtering approaches~\cite{tananti,xiong-icse18,xin2017identifying} can be further incorporated. 

\subsection{Contribution of Each Component (RQ3)}
% As introduced in Section~\ref{sec:app}, 
\tool{} incorporates two major components for patch generation, i.e., offline token-level pattern mining and online expression-level code search. To evaluate the effectiveness of each component, we have repeated our experiment by removing each component, respectively. Table~\ref{tab:component} presents the experimental results. The offline token-level pattern mining contributed 42 correct fixes and expression-level code search contributed 33 correct fixes, demonstrating that both components largely contributed to the overall effectiveness of \tool{}. However, from the table, we can also observe that the expression-level code search tends to generate more plausible but incorrect patches compared with the token-level pattern mining, which is consistent with the conclusions in prior studies~\cite{DirectFix}. Our further analysis of the results shows that most of incorrect patches are generated due to referencing infrequent code. In contrast, token-level pattern mining takes the frequency into consideration, which potentially can effectively filter out many incorrect patches. In summary, the two components in \tool{} complement each other.

In addition, as introduced in Section~\ref{sec:app}, our token-level pattern mining process incorporates a \textit{skip-fashion} pattern mining process, which can find more potential reference patterns for patch generation. Therefore, we further conducted an experiment, where we only use the token-level repair but disable the \textit{skip-fashion} in pattern mining (i.e., MAX\_SKIP=0 in Algorithm~\ref{alg:mining}). The column of ``No Skip'' in Table~\ref{tab:component} presents the results. It shows that this \textit{skip-fashion} mining process is effective since its removal caused 42-24=18 fewer bugs to be repaired. Moreover, the patch precision is also decreased from 65.6\% to 44.4\%, demonstrating the value of this process in \tool{}. Additionally, we further evaluated the performance of the S-TAC form in our expression-level repair. Specifically, we did not transform the reference code into S-TAC. Instead, they will be decomposed into the traditional TAC format without removing the code structures based on the abstract syntax tree. The experimental results are also presented in Table~\ref{tab:component} (i.e., ``No S-TAC''). After removing the S-TAC form, \tool{} failed to repair any bugs by searching reference code online. We further analyzed the results and found that the contexts of most reference code from online search are different from the faulty code, making the code adaptation fail to generate correct patches. On the contrary, 22 incorrect patches were generated. 
In summary, the experimental results demonstrate that both the \textit{skip-fashion} online pattern mining and the S-TAC form in the online search process are effective.

\begin{table}[!t]
\caption{Bugs fixed by each component (\#correct/\#plausible).}
% In the table, x/y denotes the corresponding approach generates correct patches for x bugs and generates plausible patches for y bugs.}
	\label{tab:component}
	\centering
	\resizebox{0.98\columnwidth}{!}{
% \begin{threeparttable}
\begin{tabular}{l|c|c||c|c}
\toprule
\textbf{Components} & Offline Mining & Online Search & No Skip & No S-TAC \\ \midrule
\textbf{Total}    & 42/64                         & 33/85         & 24/54            & 0/22  \\  
\textbf{Precision (\%)}    & 65.6                         & 38.8        &   44.4          & 0    
\\ \bottomrule     
\end{tabular}
% \end{threeparttable}}
}
% \vspace{-15pt}
\end{table}

\subsection{The Impact of Pattern Frequency (RQ4)}

As introduced in Section~\ref{sec:conf}, the threshold of pattern frequency is 3 by default. To investigate its impact on the effectiveness of our approach, we have conducted an additional experiment when using different frequency settings for MIN\_SUPPORT, i.e., 2, 3, and 5. As a consequence, when adopting different settings, the performance of \tool{} was not largely affected. Specifically, \tool{} correctly repaired 43/91, 42/64, and 32/68 bugs by the token-level patterns when using the threshold as 2, 3, and 5, respectively. 
In other words, our approach is not very sensitive regarding the number of correct patches when the threshold is small (i.e., 2 or 3). However,  when the threshold is too large (such as 5), \tool{} tends to repair much fewer bugs. The reason is evident as the large threshold will cause fewer patterns that can be used for patch generation. On the contrary, a smaller threshold (i.e., 2) tends to produce much more incorrect patches (i.e., 91-43=48) since the referenced patterns may be not general enough for reuse, which will affect the usability of \tool{}. 
In summary, the recall and precision of program repair indeed contradict each other. Effective patch generation techniques are always in an urgent need. In this paper, we have moved forward towards this direction. 
% \ye{When the thresholds are 2 and 3, the repair effect of \tool{} is similar. However, when the threshold is too large (such as 5), the effect will decrease. } \st{The results show that our approach is insensitive to the threshold.}\zijie{A lower frequency threshold means easier pattern generation and more bugs to fix. Nevertheless, a lower threshold results in some elements that are not actually related being recognized as patterns, which reduces the precision of fixing.}
% \z{Section 5.4 studies the sensitivity of the Repatt approach to some parameters in its algorithms; it's too ad hoc, and some statement doesn't make sense: "Repatt correctly repaired 43/91, 42/64, and 32/68 bugs...when using the threshold as 2, 3, and 5, respectively. The results show that our approach is insensitive to the threshold." The repair numbers look so different to me, why would it be claimed to be insensitive?}
% However, when it is too small (i.e., 2), although it may repair one more bug, much more plausible but incorrect patches (i.e., 91-43=48) will also be produced and thus affect the usability of \tool{}. 
In general, setting the threshold as 3 can produce relatively better results.

% To explore the impact of different frequent thresholds in the token-level fix, we execute our codebase construction fix part with $n_1$ set to be 2. As we can see from Table~\ref{tab:frequency}. Results show that the low frequent threshold leads to 5 new correct patches, meanwhile 21 new plausible patches. It also made the patch generated for Lang-7 become a plausible one.

% \input{table/frequency.tex}
\section{Discussion}
\label{sec:diss}

%Discuss the limitations (if have) and threats to validity.
%\jcom{Limitation: (1) repairing bugs that demand multiple-line code changes. (2) The existence of reference code.
%Threats to validity: (1) external threats: used benchmark for generality, 
%(2) internal threats: implementation correctness. 
%}

\textbf{Limitation:} 
% As explained, our approach is based on the ``plastic surgery hypothesis", which requires the existence of reference code. 
In this study, we have explored the feasibility of reusing finer-grained code elements for patch generation. For complex bugs that require multi-line changes, the reference code may not exist, where our approach will be less effective. In this case, our approach may potentially be combined with others since they complement each other according
to our experimental results.
% \add{Recently, many LLM-based methods have been proposed. Due to these methods require various settings when fix different bugs (e.g., temperature, top-k, beam-search and prompt format), it is challenging to make a fair comparison between \tool{} and these methods. However, it is out of scope of this study and we leave a more detailed comparison of \tool{} with these methods as our future work.}

% Recently, lots of LLM-based APR methods have been proposed. Since these methods often have high running costs and are suitable to different scenarios, we will make a more comprehensive comparison between \tool{} and these methods in our future work.

% \noindent\textbf{Threats:} 
\cradd{\textbf{Internal threats to validity.} Similar to existing works, our study also face an internal threat of manually verifying plausible patches to identify correct patches. To address this, the first three authors independently performed a careful analysis of each patch and reached a consensus. In particular, we have also made all generated patches publicly available for further inspection and validation.}

\cradd{\textbf{External threats to validity.} The primary external threat to validity lies in the subjects used in our evaluation, and the performance of \tool{} may not be generalized to other datasets. To address this, we evaluate \tool{} on two widely used datasets -- Defects4J v1.2 and v2.0, which together contain over 800 real-world bugs from 17 diverse projects. Our results demonstrate that \tool{} is effective and achieves promising performance. In the future, we plan to evaluate \tool{} on more datasets to further address this threat. Moreover, our study also share a common external threat with many existing studies -- due to the complexity and resource constraints, we reused the results of baselines reported in their papers without reproducing them in our own environment. To address this, we carefully examined the open-source repositories of the baselines and adhere to the common 5-hour time budget setting used in APR. In the future, we plan to conduct a more comprehensive evaluation of both our approach and the baselines a unified environment to further address this threat.}

\crdel{Like existing studies, we face the \textit{internal} threat of manually verifying plausible patches. To mitigate this, we rigorously compared them with developer patches for semantic equivalence and publicly released all generated patches for further check. The threats to \textit{external} validity mainly lie in the subjects used in our study. Since the Defects4J benchmark has been widely used by existing studies and contains more than 800 real-world bugs from 17 diverse projects, we believe the results can be reliable. However, the effectiveness of our approach on a broader range remains to be investigated.}
\begin{table}[ht]

\label{tab:related}
  \centering
  \caption{Comparison of Repatt with Existing Redundancy-Based APR Approaches}
  \resizebox{\columnwidth}{!}{
    \begin{tabular}{c|c|c|c}
    \toprule
    Approach & Reuse Granularity & Patch Generation Technique & Context Awareness \\
    \midrule
    GenProg & Statement & Genetic Programming (GP)-based random search & No \\
    \midrule
    AE    & Statement & Systematic search pruned by semantic equivalence	 & No \\
    \midrule
    RSRepair & Statement & Random application of mutation operators	 & No \\
    \midrule
    ARJA  & Statement & Multi-objective GP with type-matching	 & No \\
    \midrule
    ssFix & ASTNode & Syntactic search and code transfer	 & YES \\
    \midrule
    SimFix & ASTNode & Similarity search and pattern mining & YES \\
    \midrule
    CapGen & ASTNode & Context-aware search and ranking of AST node edits	 & YES \\
    \midrule
    SearchRepair & Statement & Semantic search via SMT constraint solving	 & YES \\
    \midrule
    HERCULES & ASTNode & Simultaneous repair of abstract sibling locations	 & YES \\
    \midrule
    TransplantFix & ASTNode & Graph differencing of CFGs to transplant methods	 & YES \\
    \midrule
    Repatt & Token & Pattern matching using mined token/S-TAC patterns	 & YES \\
    \bottomrule
    \end{tabular}
    }
\end{table}%

\section{Related Work}
\label{sec:related}

% \z{The related work section is not satisfying. There should be a more extensive discussion comparing Repatt with the most related search-based techniques such as SimFix, CapGen, and ssFix, showing what's in common and more importantly what's unique in Repatt and how the unique features make a difference.}
There are many automated program approaches have been proposed and achieved good results~\cite{GenProg,genesis,AE,Autofix-E,B.Le2016,DirectFix,SemFix,mechtaev2018semantic,ISSTA18-SimFix,xiong-icse17,xuan2016nopol,Xuan2016History,ghanbari2018practical,jiang2023impact,jiang2021cure}, among which redundancy-based APR approaches have been attracted much attention, such as GenProg~\cite{GenProg,GenProgTSE}, AE~\cite{AE,yuan2018arja}, RSRepair~\cite{RSRepair}, ARJA~\cite{yuan2018arja}, ssFix~\cite{ssFix}, SimFix~\cite{ISSTA18-SimFix}, CapGen~\cite{Wen2018ContextAwarePG}, SearchRpepair~\cite{Ke15ase},
HERCULES~\cite{saha2019harnessing} and so on. All these approaches are based on the ``plastic surgery hypothesis"~\cite{DBLP:conf/sigsoft/BarrBDHS14}.
% that assumes the patch ingredients can be found in either the faulty projects or some other projects. 
\cradd{Specifically, Table~\ref{tab:related} summarizes }\crdel{The}\cradd{the} major differences among them \crdel{lies in the adopted patch generation strategies}\cradd{, including code reuse granularity, patch generation strategies and whether they are context-aware}. \crdel{For example, GenProg reuses existing code randomly, ssFix searches similar code as references from a large codebase, while SimFix further incorporates history patches to refine the patch space. Similarly, CapGen considers the frequency of tokens for patch generation while HERCULES leverages the co-evolution of program elements.}\cradd{For example, GenProg reuses code statement using a genetic programming-based random search and is not context aware, ssFix retrieves syntactically similar code of AST and will rename variables to fit current context. SimFix further incorporates history patches to refine the patch space and consider the context. Similarly, CapGen considers the frequency of tokens for patch generation while HERCULES leverages the co-evolution of program elements.}
% As explained in the introduction, our approach aims to improve the performance of redundancy-based APR by exploring finer-grained code reuse, which is different from existing approaches. Specifically, compared with SimFix and ssFix, our approach leverages finer-grained token-level code patterns while SimFix and ssFix cannot as they depend on the AST difference between buggy and referenced code. Similarly, our approach is also different from CapGen as \tool{} generates patches by mining code patterns while CapGen only considers the frequency of tokens without considering their combinations. In addition, compared with SearchRepair, our approach does not rely on the heavy semantic code search process. In particular, our approach can leverage nonconsecutive token usage patterns for patch generation, which has been proved to be effective but is not supported by these existing approaches.
Our approach enhances redundancy-based APR by enabling finer-grained code reuse, differentiating it from existing methods. Unlike SimFix and ssFix, which rely on AST differences, our method operates at the token level, capturing more granular code patterns. Compared to CapGen, which considers token frequency without their contextual relationships, our approach mines structured code patterns for patch generation. Additionally, unlike SearchRepair, we avoid costly semantic code searches and uniquely support nonconsecutive token usage patterns, which have been shown to be effective but are not utilized by existing methods.
% Additionally, PAR~\cite{PAR} and TBar~\cite{liu2019tbar} aim to improve the patch quality via pre-defining a set of repair templates.
% Apart from those, a set of APR techniques leverage the advance of mature constraint-solving techniques for patch generation~\cite{SemFix,DirectFix,Mechtaev,mechtaev2018semantic,gao2021beyond,xuan2016nopol,JAID,SKETCHFIX}. However, since these approaches depend on symbolic execution and constraint solving, they usually face the scalability issue for practical use.
% while our approach does not.
Other APR techniques, such as PAR~\cite{PAR} and TBar~\cite{liu2019tbar}, improve patch quality through predefined repair templates, while constraint-solving-based techniques~\cite{SemFix,DirectFix,Mechtaev,mechtaev2018semantic,gao2021beyond,xuan2016nopol,JAID,SKETCHFIX} rely on symbolic execution and constraint solving, often facing scalability challenges.
Different from these approaches, this work aims to improve the performance of redundancy-based APR techniques and does not face the scalability issue.

With the rapid development of deep learning techniques, the latest APR approaches also leverage such techniques for patch generation. Early studies use statistical machine learning algorithms to sort or pick patch ingredients. For example, Prophet~\cite{long2016automatic}, ACS~\cite{xiong-icse17}, Elixir~\cite{ELIXIR}, Hanabi~\cite{xiong2022l2s}, and LIANA~\cite{chen2022program} all use different learning models for patch ingredient selection. More recently, state-of-the-art deep learning techniques have been employed in APR techniques, such as SEQUENCER~\cite{chen2019sequencer}, CoCoNut~\cite{lutellier2020coconut}, CURE~\cite{jiang2021cure}, DLFix~\cite{li2020dlfix}, Recoder~\cite{zhu2021syntax}, RewardRepair~\cite{ye2022neural}, SelfAPR~\cite{ye2022selfapr}, and many others~\cite{fan2022automated,xia2023automated}. These approaches suffer from interpretability issues and mostly can repair simple bugs (e.g., single-line bugs). In contrast, our approach generates patches by reusing existing code, which complements them based on our experimental results. The latest LLM-based APR methods further improved repair performance~\cite{xia2022less,prenner2022can, xia2023automated, jiang2023impact, li2024hybrid}. 
% \add{Recently, many works employ LLMs for APR, treating program repair as a task of code generation. AlphaRepair~\cite{xia2022less} pioneered LLM-based infilling-style APR by masking buggy code and using CodeBERT~\cite{feng2020codebert} to replace masked tokens with correct ones. Several studies~\cite{prenner2022can, xia2023automated, jiang2023impact} have further explored applying various LLMs to APR, highlighting the potential of LLMs in APR. To better apply LLMs to APR tasks, 
For example, Repilot~\cite{wei2023copiloting} integrates CodeT5~\cite{wang2021codet5} with a completion engine to enhance repair performance, while GAMMA~\cite{zhang2023gamma} utilizes CodeBERT and UniXcoder~\cite{guo2022unixcoder} to fill predefined repair templates. Unlike \tool{}, these methods treat LLMs as black boxes and are still facing challenges related to interpretability~\cite{zhang2024systematic}, data leakage~\cite{sallou2024breaking}, and generalizability~\cite{li2025evaluating}. In addition, according to our evaluation results, our approach also complements these LLM-based methods, and thus can be further combined with them for better APR.

% \subsection{Semantic-based APR}
% Semantic-based APR approaches reduce patch generation to a constraint-solving problem by extracting correctness constraints from tests and code structure~\cite{SemFix,DirectFix,Mechtaev,mechtaev2018semantic,gao2021beyond,xuan2016nopol,JAID,SKETCHFIX}. These approaches either fully or partially execute the buggy program using symbolic execution, or summarize the behavior of tests to collect constraints. However, symbolic execution and constraint summarization have limitations, such as the path explosion problem and the ability of SMT-solvers. Our approach can avoid these limitations.

% \subsection{Data Mining for Source Code Patterns}
% Data mining techniques have been widely applied to source code analysis for various tasks such as code search, code clone detection, code correctness judgment, and code-convention mining~\cite{le2020deep,wu2022detecting,wang2020automated,lin2022context,tian2022best,allamanis2014learning}. Source code exhibits naturalness, and discovering and applying code patterns is an effective way to analyze source code~\cite{Hindle2012,ray2016naturalness}. Our approach to APR employs a novel mining policy at a finer-grained level.
\section{Conclusion}
\label{sec:conclude}
% Redundancy-based APR techniques have been widely studied and many approaches have been proposed.
% However, they often generate a large number of incorrect patches while only effectively repairing a small number of the bugs. 
To enhance redundancy-based APR, this paper proposes \tool{}, a novel repair technique using a two-level pattern mining process for precise patch generation via fine-grained code reuse. Evaluations on Defects4J v1.2 and v2.0 show that \tool{} complements existing methods by repairing unique bugs and improving patch precision in real-world repair scenarios. Furthermore, we present the first attempt at combining 
different approaches to improve the SOTA APRs. 
Our promising results encourage further research, and we have open-sourced our data and implementations for replication and exploration~\cite{homepage}.
 % Additionally, we pioneer the combination of different approaches to advance SoTA APR. Our promising results encourage further research, and we have open-sourced our data and implementations for replication and exploration~\cite{homepage}.
% \hy{revise a bit, pls see if it is ok}

% However, they can still repair a small number of bugs with generating a large number of incorrect patches. Targeting to improve the performance of redundancy-based APRs, this paper proposes a new repair technique, named \tool{}, which constitutes a two-level pattern mining process for more effective patch generation through fine-grained code reuse. The evaluation on the Defects4J v1.2 and v2.0 benchmarks shows that \tool{} complements existing approaches by repairing many unique bugs and achieves higher patch precision under the real repair practice. Additionally, we also made the first attempt to combine different approaches for improving the state-of-the-art APRs. 
% The promising results motivate more studies in this direction.

% \section*{Data Availability}
% We make all our implementations and experimental results publicly available at:
% \plink{}.

\section*{Acknowledgment}
We thank the editors and anonymous reviewers for their constructive suggestions to help improve the quality of this paper. This work was supported by the National Key Research and Development
Program of China (Grant No. 2024YFB4506300), and the National Natural Science Foundation of China (Grant Nos. 62202324, No. 62322208 and No. 62202040).
% \end{acks}

\balance
% \clearpage 
% \input{ref.bib}

\bibliographystyle{IEEEtran}

\end{document}